\documentclass[]{aastex631}

\usepackage{lineno}

\usepackage{ulem}
\usepackage{balance}

\newcommand\kms{{\rm\,km\,s^{-1}}}

 \def\mso{\,M_\odot}

 \def\simle{\mathrel{\hbox{\rlap{\hbox{\lower4pt\hbox{$\sim$}}}\hbox{$<$}}}}
 \def\simgr{\mathrel{\hbox{\rlap{\hbox{\lower4pt\hbox{$\sim$}}}\hbox{$>$}}}}
\renewcommand{\thefigure}{\arabic{figure}}

\received{2024 August 14}
\revised{2024 October 1}
\accepted{2024 October 14}
\submitjournal{ApJL}

\shorttitle{Stripped helium-star and compact object binaries in coeval populations}
\shortauthors{Chen Wang et al.}

\begin{document}

\title{Stripped helium-star and compact object binaries in coeval populations \\--- predictions based on detailed binary evolution models}

\correspondingauthor{Chen Wang}
\email{cwang@mpa-garching.mpg.de}

\author[0000-0002-0716-3801]{Chen Wang}
\affiliation{Max Planck Institute for Astrophysics, Karl-Schwarzschild-Strasse 1, 85748 Garching, Germany}

\author[0000-0002-9552-7010]{Julia Bodensteiner}
\affiliation{European Southern Observatory, Karl-Schwarzschild-Str. 2 85738 Garching bei München, Germany}

\author[0000-0001-9565-9462]{Xiao-Tian Xu}
\affiliation{Argelander-Institut f\"ur Astronomie, Universit\"at Bonn, Auf dem H\"ugel 71, 53121 Bonn, Germany}

\author[0000-0001-9336-2825]{Selma E. de Mink}
\affiliation{Max Planck Institute for Astrophysics, Karl-Schwarzschild-Strasse 1, 85748 Garching, Germany}

\author[0000-0003-3026-0367]{Norbert Langer}
\affiliation{Argelander-Institut f\"ur Astronomie, Universit\"at Bonn, Auf dem H\"ugel 71, 53121 Bonn, Germany}

\author[0000-0003-1009-5691]{Eva Laplace}
\affiliation{Heidelberger Institut für Theoretische Studien, Schloss-Wolfsbrunnenweg 35, D-69118 Heidelberg, Germany}

\author[0000-0003-1817-3586]{Alejandro Vigna-G\'omez}
\affiliation{Max Planck Institute for Astrophysics, Karl-Schwarzschild-Strasse 1, 85748 Garching, Germany}

\author{Stephen Justham}
\affiliation{Max Planck Institute for Astrophysics, Karl-Schwarzschild-Strasse 1, 85748 Garching, Germany}

\author[0000-0002-7527-5741]{Jakub Klencki}
\affiliation{European Southern Observatory, Karl-Schwarzschild-Str. 2 85738 Garching bei München, Germany}

\author[0000-0002-6105-6492]{Aleksandra Olejak}
\affiliation{Max Planck Institute for Astrophysics, Karl-Schwarzschild-Strasse 1, 85748 Garching, Germany}

\author[0000-0003-3456-3349]{Ruggero Valli}
\affiliation{Max Planck Institute for Astrophysics, Karl-Schwarzschild-Strasse 1, 85748 Garching, Germany}

\author[0000-0002-2715-7484]{Abel Schootemeijer}
\affiliation{Argelander-Institut f\"ur Astronomie, Universit\"at Bonn, Auf dem H\"ugel 71, 53121 Bonn, Germany}

\begin{abstract}
Massive stars mainly form in close binaries, where their mutual interactions can profoundly alter their evolutionary paths. Evolved binaries consisting of a massive OB-type main-sequence star with a stripped helium star or a compact companion represent a crucial stage in the evolution towards double compact objects, whose mergers are (potentially) detectable via gravitational waves. The recent detection of X-ray quiet OB+black hole binaries and OB+stripped helium star binaries has set the stage for discovering more of these systems in the near future. In this work, based on 3670 detailed binary-evolution models and using empirical distributions of initial binary parameters, we compute the expected population of such evolved massive binaries in coeval stellar populations, including stars in star clusters and in galaxies with starburst activities, for ages up to 100\,Myr.
Our results are vividly illustrated in an animation that shows the evolution of these binaries in the color-magnitude diagram over time. We find that the number of OB+black hole binaries peaks around 10\,Myr, and OB+neutron star binaries are most abundant at approximately 20\,Myr. Both black holes and neutron stars can potentially be found in populations with ages up to 90\,Myr. Additionally, we analyze the properties of such binaries at specific ages. We find that OB+helium stars and OB+black hole binaries are likely to be identifiable as single-lined spectroscopic binaries. Our research serves as a guide for future observational efforts to discover such binaries in young star clusters and starburst environments.

\end{abstract}

\section{Introduction} \label{sec:intro}
The first direct detection of gravitational waves (GWs) in 2015 marked a new era in our exploration of the Universe \citep{2016PhRvL.116f1102A}. Since then, nearly a hundred events have been observed by the LIGO/Virgo/KAGRA collaboration \citep{2023PhRvX..13d1039A}. One of the most important sources of these GWs are emitted when two double compact objects merge. These compact objects are the end products of massive star evolution. Several formation scenarios for these double compact objects have been proposed, including dynamical processes \citep{1993Natur.364..421K,2010MNRAS.402..371B,2016ApJ...816...65A,2019MNRAS.487.2947D} and isolated binary evolution \citep{1993MNRAS.260..675T,2007PhR...442...75K,2016Natur.534..512B,2018MNRAS.481.1908K}. We refer to \cite{2022PhR...955....1M} for a review.

Studies on the multiplicity of stars indicate that 70\% of massive stars will interact in close binaries \citep{2012Sci...337..444S,2017ApJS..230...15M}. However, accurately predicting their evolutionary outcomes remains a challenge (see \citealp{2012ARA&A..50..107L} and \citealp{2023arXiv231101865M} for a review). Binary evolution encompasses several poorly understood stages \citep{2007A&A...467.1181D,2016MNRAS.456..485S,2022ARA&A..60..455E,2023arXiv231101865M,2024A&A...682A.169H}, which can lead to substantial variations in predicted frequencies and characteristics of their final outcomes \citep{2022ApJ...925...69B,2021A&A...647A.153B,2021ApJ...910..152Z,2021A&A...651A.100O,2022ApJ...940..184V,2024A&A...685A.169D}.

During the course of binary evolution, phases involving a stripped helium star (HeS) or a compact companion (cc) in conjunction with a main-sequence (MS) OB star are of significant interest. These stages represent the last long-lived stages that can be traced from the double MS stage through detailed stellar evolution calculations. Beyond this stage, binaries are likely to enter the short-lived common envelope evolution, whose outcome is highly uncertain and generally oversimplified \citep{2013A&ARv..21...59I,2020cee..book.....I,2023LRCA....9....2R}. Moreover, these binaries are anticipated to be quite common, as a substantial proportion of binaries are capable of avoiding a merger in the first mass transfer phase and evolve to such systems  \citep{2014ApJ...796...37S,2019ApJ...885..151S,2020A&A...638A..39L}.

However, detecting such binaries is challenging, as stripped helium stars and compact objects are typically optically faint \citep{2017A&A...608A..11G}. While X-ray observations have identified OB+neutron star (NS) binaries \citep[see][for a review]{2011Ap&SS.332....1R}, the majority of OB+black hole (BH) binaries may be X-ray quiet \citep{2020A&A...638A..39L,2021A&A...652A.138S,2021PASA...38...56H}, known as inert BH binaries. For instance, \citet{2020A&A...638A..39L} predicted more than 100 OB+BH binaries in the Large Magellanic Cloud (LMC) based on detailed binary evolution models, but only LMC X-1 had been confirmed via X-rays \citep{2009ApJ...697..573O}. Nevertheless, recent spectroscopic observations have identified X-ray quiet OB+BH binaries, including VFTS 243 \citep{2022NatAs...6.1085S} and two candidates, VFTS 514 and VFTS 779 in the LMC \citep{2022A&A...665A.148S}. Other confirmed inert BH systems include HD\,130298 \citep{2022A&A...664A.159M}, and Gaia BH1, Gaia BH2 and Gaia BH3 \citep{2023MNRAS.521.4323E,2023MNRAS.518.1057E,2023MNRAS.521.4323E,2024A&A...686L...2G}, with HD\,130298 paired with an O-type companion, and the others with solar-like stars.

Binaries with intermediate-mass HeS ($1-8\mso$), have only recently been confirmed through UV photometry \citep{2023Sci...382.1287D,2023ApJ...959..125G}. Additionally, OB star plus partially-stripped star binaries have been discovered, in which the partially-stripped stars are still bloated B-type stars rapidly contracting towards the subdwarf regime \citep{2022A&A...667A.122S,2023A&A...674L..12R,2024arXiv240100802Y,2024arXiv240617678R}. Notable examples of such binaries include LB-1 \citep{2020A&A...633L...5I,2020A&A...639L...6S}, HR\,6819 \citep{2020A&A...641A..43B,2021MNRAS.502.3436E} and VFTS\,291 \citep{2023MNRAS.525.5121V}.

Recent discoveries show that previously significant challenges are now attainable through high-quality observations, paving the way for future findings. It is increasingly important to refine theoretical predictions for evolved massive binaries using state-of-the-art models. Prior research has often relied on rapid binary evolution models (e.g. \citealt{2002MNRAS.329..897H} and their derivatives such as \citealt{2004MNRAS.348.1215I,2018MNRAS.481.1908K,2020A&A...636A.104B}). These studies apply rough approximations to model the mass transfer process, and primarily focused on reproducing populations of X-ray binaries and GW sources \citep{2000A&A...358..462V,2006csxs.book..623T,2023pbse.book.....T,2016MNRAS.462.3302E,Pablo2016,2018MNRAS.481.4009V}. There remains a significant gap in theoretical studies for OB+HeS/cc binaries. Existing works employing detailed binaries models for such binaries have focused on field stars \citep{2020A&A...638A..39L, 2022A&A...667A..58P, 2024Xu}. However, selecting suitable targets for multi-epoch spectroscopic observations among field stars is challenging, as it is difficult to ascertain whether a field star has experienced binary interaction.

In this context, post-interaction systems in coeval stellar populations may stand out \citep{1994A&A...288..475P,1998A&A...334...21V,2011ApJ...731L..37Y,2020ApJ...888L..12W}, in particular, occupying distinct positions in the color-magnitude diagram (CMD).
Young open clusters are the most important sources of coeval stellar populations. They are ideal for studying stellar evolution, as the influence of dynamical interactions, which plays a significant role in older globular clusters, is much less important in young open clusters. Additionally, starburst galaxies, common in the distant Universe \citep{2015MNRAS.452.1447K,2019ApJ...876....3L}, also provide coeval stellar populations for study. For example, recent star formation peaks at approximately 12 and 10\,Myr have been identified for the LMC and Small Magellanic Clouds (SMC), respectively \citep{2009AJ....138.1243H,2018MNRAS.478.5017R}. Among galaxies with starburst activities, galaxies that show distinct spectral signatures of hot and luminous Wolf-Rayet (WR) stars are of particular interest \citep{2008A&A...485..657B,2024arXiv240410037M}, as they play a significant role in the ionization of the early universe through their high-energy radiation \citep{1991ApJ...377..115C}.
Given that recent advancements in spectroscopic observations enable the characterization of individual stars in young open clusters \citep{2020A&A...634A..51B,2021A&A...652A..70B,2023A&A...680A..32B,2020MNRAS.492.2177K,2023MNRAS.518.1505K,2023MNRAS.526..299S}, and 
the active interest in searching for post binary interaction systems, investigating coeval population of post-interaction systems is timely and will provide valuable guidance for observers in effectively pinpointing these systems.

In this study, we utilize a population of detailed binary models to examine the number and characteristics of OB+stripped HeS/cc binaries (referred to as evolved massive binaries in this study) in coeval stellar populations up to 100\,Myr. 
The paper is structured as follows: Section\,\ref{sec:method} provides a brief introduction to our models and the underlying physics assumptions. In Section\,\ref{sec:Results}, we present our predictions for evolved massive binaries. We compare our results with stars in young open clusters in Section\,\ref{sec:compare_ob}. Section\,\ref{sec:discussion} discusses the impact of uncertain physics on our results and Section\,\ref{sec:conclusion} summarizes our conclusions.

\section{Detailed binary evolution models}\label{sec:method}
\subsection{Initial binary parameters}
In this work, we analyze the detailed binary evolution models in \citet{2022NatAs...6..480W}, whose initial parameters are created from Monte Carlo sampling. The initial primary masses are from 3 to $100\mso$ following the \cite{Salpeter1955} initial mass function (IMF). The initial mass ratios span from 0.1 to 1 \citep{2012Sci...337..444S,2022A&A...665A.148S}, adhering to a uniform distribution. The initial orbital periods range from the minimum value, where the two stellar components are initially in contact at their surfaces, to 8.6\,yrs ($\log P_{\rm i}\, /\rm d=3.5$), assuming a flat distribution in logarithmic space \citep{1924PTarO..25f...1O,1983ARA&A..21..343A,2023A&A...674A..60B}.

The upper mass limit of young open clusters in the Magellanic Clouds is on the order of $10^5 \mso$ \citep{2003AJ....126.1836H,2005ApJS..161..304M,2012ApJ...751..122P}. 
To reflect this, we modeled the evolution of 3670 detailed binary systems, using the initial parameter distributions mentioned earlier. The total initial mass of these binaries is approximately $4.7\times 10^4\mso$. Assuming a binary fraction of 100\%, a lower mass limit of $0.08\mso$ and the Kroupa IMF with a power-law index of 1.3 for stars between $0.08\mso$ and $0.5\mso$ \citep{2001MNRAS.322..231K}, our simulations represent a cluster with a total stellar mass of approximately $1.3 \times 10^5 \mso$.

These simulations cover a broad range of initial parameters and mass transfer scenarios, including stable Case A, Case B \citep{1967ZA.....65..251K}, and unstable mass transfer. Our goal is to explore the formation mechanisms of evolved massive binaries in young open clusters. We address the effects of finite sample size in Sec.\,\ref{sec:discussion_size}, and our results can be upscaled for starburst galaxies.

We assume both binary components rotate at 55\% of their critical values at their zero-age MS (ZAMS), based on the estimated rotation rates for the majority of MS stars in young open clusters \citep{2019ApJ...887..199G,2022NatAs...6..480W}. Comparing with non-rotating detailed binary models in \cite{2020ApJ...888L..12W}, we found that intermediate rapid rotation does not significantly alter the overall binary evolutionary outcomes. However, stars with initially intermediate rapid rotation can approach near critical rotation at the end of their MS evolution through single star evolution \citep{2008A&A...478..467E,2020A&A...633A.165H}. This means that the initial rotation can impact the predicted number of near critically-rotating Be stars near the cluster turn off.

We adopt SMC-like metallicity ($Z_{\rm SMC}=0.002$), using the same chemical compositions as described in \cite{2011A&A...530A.115B}. Choosing a low-metallicity environment reduces uncertainties from stellar winds, allowing us to attribute any exotic characteristics in our binary models mainly to binary interactions. 

\subsection{MESA input physics}
The binary evolution is computed using the one-dimensional stellar evolution code Modules for Experiments in Stellar Astrophysics (MESA, version 8845, \citealt{Paxton2011,Paxton2013,Paxton2015}), with detailed physics assumptions from \citet{2022NatAs...6..480W}. Here we only briefly summarize the relevant physics assumptions. 

We consider the standard mixing-length theory with a mixing-length parameter of $\alpha=l/H_{P}=1.5$, where $l$ and $H_{P}$ are the mixing length and the local pressure scale height, respectively \citep{2015A&A...580A..20S,2017A&A...597A..71S}. We apply the Ledoux criterion to determine the boundaries of convective zones. We adopt a mass-dependent step-overshooting parameter, which can better explain the observed trend that the width of the distribution of field MS stars in the HRD increases with stellar mass \citep{2014A&A...570L..13C,2016A&A...592A..15C,2019A&A...625A.132S,2021A&A...653A.144H}. The value $\alpha_{\rm OV}=0.3$ is used for models with initial masses larger than $20\mso$ \citep{2011A&A...530A.115B}. Below this mass, $\alpha_{\rm OV}$ decreases linearly such that it equals 0.1 at an initial mass of $1.66\mso$. Below $1.66\mso$, $\alpha_{\rm OV}$ drops quickly such that it reaches 0 at an initial mass of $1.3\mso$ \citep{2016A&A...592A..15C}. Semiconvective mixing uses $\alpha_\mathrm{SC}=1$ \citep{1983A&A...126..207L}. Rotational mixing accounts for various instabilities, including dynamical and secular shear instabilities, the Goldreich-Schubert-Fricke instability, and Eddington-Sweet circulations \citep{2000ApJ...528..368H}, with a mixing efficiency of $f_\mathrm{C}=1/30$ \citep{1992A&A...253..173C}. The Tayler-Spruit dynamo is included for angular momentum transport \citep{Spruit2002,Heger2005,2019MNRAS.485.3661F}.

Mass loss follows \citet{2011A&A...530A.115B}. For stars with surface hydrogen mass fraction $X_\mathrm{S}\geq0.7$, we utilize the wind prescription of \cite{2001A&A...369..574V}. For stars with $X_\mathrm{S}\leq0.4$, we adopt the WR mass-loss recipe of \cite{Hamann1995}, scaled down by a factor of 10 to account for clumping effects \citep{2006A&A...460..199Y,2011A&A...530A.115B}. For stars with $0.4<X_\mathrm{S}<0.7$, a linear interpolation of the logarithmic mass-loss rate $\log\, \dot{M}$ between the two prescriptions is adopted. The metallicity-dependent stellar winds scales as $\dot{M} \propto Z^{0.85}$ \citep{2001A&A...369..574V}. 

Recent studies have indicated that the mass-loss rates for intermediate-mass (1--8$\mso$) HeS might be lower than previously estimated using the WR mass-loss schemes \citep{2016A&A...593A.101K,2017A&A...607L...8V}. Specifically, \citet{2023ApJ...959..125G} suggested that stripped HeS mass-loss rates are about an order of magnitude lower than predicted by \citet{2000A&A...360..227N} and lower than \citet{2017A&A...607L...8V}. \citet{2023MNRAS.518..860G} pointed out that the mass loss rates of \cite{Hamann1995} divided by a factor of 10, i.e. the one used in our simulations, are slightly below the metallicity-scaled \citet{2000A&A...360..227N} mass-loss rate, which is approximately one-third of the original \citet{2000A&A...360..227N} rate at SMC metallicity. This means that the order of magnitude of the mass loss rates adopted in our models is comparable to those found in \citet{2023ApJ...959..125G}. Reducing mass-loss rates, as shown by \citet{2006A&A...452..295E}, does not significantly impact the lifetime or the final mass of intermediate-mass HeSs due to their already low wind rates (on the order of $10^{-10}$ to $10^{-9}\, M_\odot\,{\rm yr^{-1}}$). Consequently, we do not expect the uncertainty in HeS mass-loss rates would significantly impact our results. 

The detection of apparently single WR stars in the low-metallicity environment of the SMC suggests that stars more massive than around $40\mso$ may undergo eruptive mass loss to shed their hydrogen envelopes in the luminous blue variable (LBV) phase \citep{2024arXiv240601420S}. This eruptive mass loss is not included in our models, and may affect the results for populations younger than approximately 6\,Myr.

We start the evolution from the ZAMS with circular orbits. The detailed structure of both binary components and the orbital evolution is considered, accounting for tidally induced spin-orbit coupling \citep{2008A&A...484..831D} and Roche-lobe overflow.
During Roche-lobe overflow, the specific angular momentum accreted by the secondary star depends on whether the accretion occurs ballistically or through a Keplerian disk \citep{2013ApJ...764..166D}. We employ rotationally enhanced stellar winds to prevent models from exceeding critical rotation \citep{1998A&A...329..551L,2005A&A...435.1013P,2012ARA&A..50..107L,Paxton2015}. We assume that any material that cannot be accreted by the mass gainer is lost from its surface, driven by the combined radiation energy of the two stars. 

In cases where the combined radiation energy is insufficient to drive the excess material to infinity, we assume that the two stars become engulfed in the excess material, leading to a binary merger. Additionally, a binary merger is also assumed to occur when mass outflow happens through the second Lagrangian point or when the mass transfer rate reaches an ad-hoc upper limit of 
$\dot{M}>10^{-1}\,M_\odot\,\mathrm{yr^{-1}}$, indicating potentially unstable mass transfer and common envelope evolution. Furthermore, under very specific conditions (extremely close binaries with high mass ratios), the secondary star can evolve more rapidly than the primary, triggering inverse mass transfer. We assume the two stars merge at the onset of inverse mass transfer due to their extreme mass ratios. This usually occurs when the secondary star finishes central hydrogen burning. The evolution of merger products of two MS stars is calculated according to \citet{2016MNRAS.457.2355S}.
The evolution of merger products of binaries containing a post-MS star is beyond the scope of this paper. If none of the aforementioned merger criteria are met, we track the evolution of the binary models until core carbon depletion occurs in both components. 

If the core mass of the primary star exceeds the Chandrasekhar limit when reaching carbon depletion, a supernova (SN) explosion is assumed, and the companion is modeled as an isolated star. The classification of SN explosion remnant is described in Sec.\,\ref{sec:assumption_SN}. As mentioned above, we do not model the reverse mass transfer from the OB star to the compact object, as this mainly happens when the OB star has finished its MS evolution, which will not affect the results in this study. 
We limit the calculations to the point of central helium exhaustion if the helium core mass exceeds $13\,M_\odot$ to maintain computational stability.

\subsection{Treatment of compact object binaries}\label{sec:assumption_SN}
Previous studies have revealed that the formation of BHs and NSs is likely non-monotonic with mass \citep{1996ApJ...457..834T,2011ApJ...730...70O,2014ApJ...783...10S,2018ApJ...860...93S,2020ApJ...890...43C,2021A&A...645A...5S} due to the effect of neutrino-dominated nuclear burning in the late evolution of massive stars \citep{2024arXiv240902058L}. However, we follow simpler assumptions from \citet{2020A&A...638A..39L} and \citet{2024Xu} that a BH is formed if the final He core mass of our models exceeds $6.6\mso$ \citep{2018ApJ...860...93S,2020A&A...638A..39L}. We may therefore overestimates the number of OB+BH binaries. We discuss this in Sections.\,\ref{sec:discussion_SN} and Appendix\,\ref{app_sec:SN}.
The mass of the BH is calculated using the same assumptions as in the ComBineE code \citep{2018MNRAS.481.1908K}. It is assumed that 20\% of the mass of the He-rich envelope is ejected, and after that 20\% of the remaining mass is lost due to the release of the gravitational binding energy. 

Models with a helium core mass below $6.6\mso$ and a CO core mass above $1.37\mso$ \citep{1984ApJ...277..791N,2018MNRAS.481.1908K} are assumed to form NSs with a mass of $1.4\mso$. Among these, the models with final CO core masses between $1.37$ and $1.435\mso$ are assumed to form NSs caused by electron-capture supernovae (ECSN), as described by \cite{1984ApJ...277..791N,1987ApJ...322..206N,2008ApJ...675..614P} and \cite{2018MNRAS.481.1908K}, for which we assume smaller kicks (see below). Models with CO core masses lower than $1.37\mso$ are assumed to form white dwarfs (WDs). 

We assume zero kicks for BHs \citep{2013MNRAS.434.1355J,2020MNRAS.495.3751C,2024PhRvL.132s1403V}. For NSs, we assume a Maxwell-Boltzmann distribution with a root-mean-square velocity $\sigma=265\kms$ \citep{2005MNRAS.360..974H} for normal core-collapse SN (CCSN) and a flat distribution between 0 and 50$\kms$ for ECSN \citep{2002ApJ...574..364P,2004ApJ...612.1044P}. The aforementioned kicks are referred to as `fiducial kicks' in this study. This is particularly important for the study of young open clusters, as OB stars might be ejected from the cluster. The post-kick velocities of these OB stars are on the order of tens of kilometers per second, which may exceed the escape velocity (typically $12-15 \kms$) of a young open cluster \citep{2014ApJ...797...35G,2010ARA&A..48..431P}. The overabundance of rapid rotators detected among runaway O-type stars suggests that these stars are post-interaction stars, ejected by their companions \citep{2018A&A...616A.149M,2022A&A...668L...5S}. 
We also model NSs with no kicks to understand OB+NS populations in starburst galaxies, where stars may not escape their host galaxies. 


\subsection{Constructing the color-magnitude diagram}
To construct the stellar distribution in the CMD, we adopt the same method as in \cite{2022NatAs...6..480W}. We convert stellar temperature and luminosity to magnitudes in the Hubble Space Telescope (HST)/WFC3 F814W and F336W bands. We consider a distance modulus of 18.82 and a reddening E(B-V) of 0.11, with which our models best match the observations of stars in the SMC cluster NGC\,330.

\section{Results}\label{sec:Results}
\subsection{Evolved massive binaries at a specific age}

\begin{figure*}[htbp]
\centering
\includegraphics[width=1\linewidth]{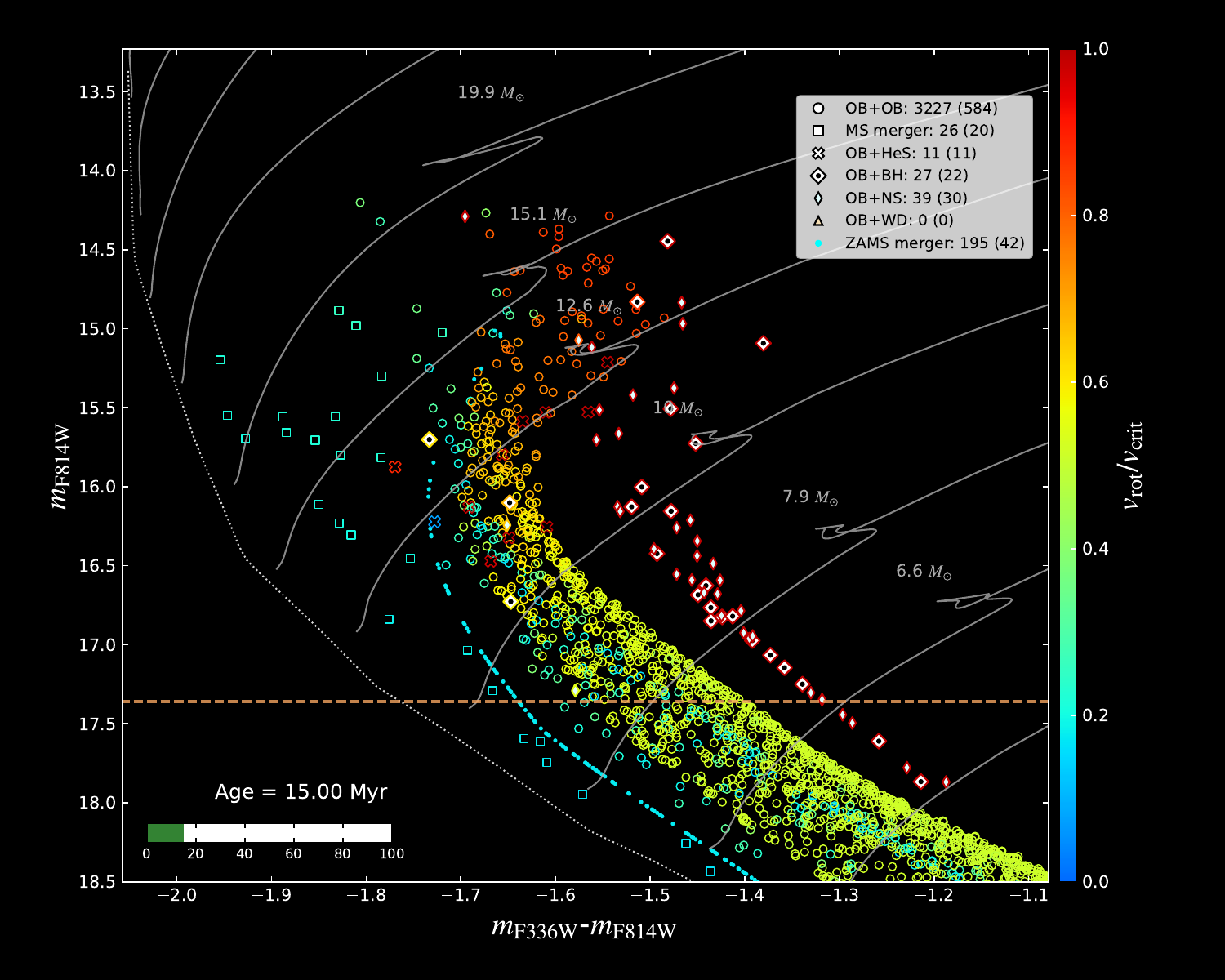}
\caption{Distribution of our models in the color-magnitude diagram at 15\,Myr, matching the estimated age of NGC\,2100. Each symbol indicates a binary system or a binary merger product, color-coded according to the rotation of the visually brighter component. We classify our models into different groups: detached binaries containing two OB main-sequence stars (open circles), main-sequence merger products (open squares), OB+helium star binaries (open crosses), OB+black hole binaries (thick diamonds with dots inside), OB+neutron star binaries (thin diamonds), OB+white dwarf binaries (triangles), and merger products of two zero-age main-sequence stars (filled dots).
For OB stars rotating at velocities higher than 95\% of their critical values, we reduce the red magnitude, $m_{\rm F814W}$, by 0.1\,mag to account for potential decretion disk flux \citep{2017AJ....153..252L}.
Grey solid curves depict evolutionary tracks of single star models with initial rotational velocities of 55\% of their critical values, and the grey dotted line marks non-rotating zero-age main-sequence positions. The orange dashed line indicates the magnitude 2.5\,mags below the cluster turn-off. The numbers outside and inside the parentheses in the legend indicate the total number of simulated models and the number of models above the orange dashed line, respectively, for each stellar population. Here we adopt a distance modulus of 18.82, representative for the SMC cluster NGC\,330 \citep{2022NatAs...6..480W}. This figure captures a snapshot from an animation depicting the distribution of our models from 2 to 100\,Myr. The full animation, with a duration of 27\,s, is available for download from the online journal.
}
\label{fig:xxu_animation}
\end{figure*}

Here, we show the properties of our evolved massive binaries at an age of 15\,Myr as a detailed example. This age aligns with the estimated age of the LMC cluster NGC\,2100 \citep{2010ARA&A..48..431P,2019MNRAS.486..266B}. At this age, our binary models predict a substantial number of binaries containing a BH or a NS. We display the distribution of our binary evolution models in the CMD in Fig.\,\ref{fig:xxu_animation}\footnote{The MS hooks of fast-rotating single star models (grey solid lines in this figure) differ from those of non-rotating star models due to the approach of critical rotation. In fast-rotating stars, nearing critical rotation during the hook phase leads to significant deformation and overall expansion. The morphology of the MS hook is shaped by the interplay between stellar contraction, driven by the depletion of nuclear fuel, and expansion, driven by stellar rotation.} and the distribution of their observable properties, including their magnitudes, semi-amplitude orbital velocities, surface rotational velocities and mass ratios, in Fig.\,\ref{fig:BH_MS_15}. The initial binary parameters for these evolved massive binaries are shown in the appendix in Fig.\,\ref{app_fig:init_par}.

Fig.\,\ref{fig:xxu_animation} reveals that post-interaction binaries are predominantly found within 3-4 magnitudes below the cluster's turn off (see also the first panel in Fig.\,\ref{fig:BH_MS_15}). As explained in \cite{2020ApJ...888L..12W}, the higher magnitude (i.e. lower luminosity) limit of these binaries in a cluster is set by the least massive secondary star that has successfully experienced stable mass transfer in the systems with an initial primary mass close to the turn-off mass of the cluster. 

Fig.\,\ref{fig:xxu_animation} also shows that the majority of OB+cc binaries are comprised of near-critically rotating OB stars, i.e. OBe stars (see also the third panel in Fig.\,\ref{fig:BH_MS_15}). These OBe+cc binaries appear significantly redder than the other coeval populations. Centrifugal forces reduce the effective gravity of fast-rotating stars, thereby reducing their effective temperature according to the von Zeipel theorem \citep{Vonzeipel1924}. The positions of these OBe+cc binaries in the CMD are somewhat uncertain, as the contributions from the disks of OBe stars to stellar color and magnitude are not precisely known. Here, we have added a value of 0.1 magnitude to OBe stars in the F814W band \citep{2017AJ....153..252L,2021A&A...653A.144H}. This 0.1 magnitude adjustment allows our model predictions to match the observed Be stars in the SMC cluster NGC\,330 \citep{2022NatAs...6..480W}. It’s important to note that even before applying the 0.1 magnitude adjustment, while the faint OBe+cc binaries ($m_\mathrm{F814W}\geq 16.5$) overlap with the reddest normal binaries, the bright OBe+cc binaries ($m_\mathrm{F814W}< 16.5$) near the cluster turn-off still appear redder than normal binaries.

OBe+cc binaries in our simulation predominantly arise from systems that have undergone a stable Case B mass transfer \citep{2020ApJ...888L..12W}, during which the OB accretor rapidly approaches critical rotation \citep{1981A&A...102...17P}, and the binary's wide orbit prevents tidal spin-down \citep{2012ARA&A..50..107L}. 
These OBe+cc binaries may be observed as X-ray binaries as the compact object passes through the decretion disk of the OBe star \citep{2016A&A...586A..81H}. OB+cc binaries containing non-critically rotating stars originate either from very wide binaries, where the two stars avoid interaction, or from very close binaries that have undergone Case A mass transfer (i.e. mass transfer that occurs when the primary star is burning hydrogen in the core). In these cases, the OB stars are either not spun-up at all or are spun-down by tidal forces. While such binaries with NS companions may be observed as X-ray binaries when the NS passes the periastron of the OB star in a high-eccentricity orbit caused by SN kicks, those with BH companions in circular orbits are usually X-ray quiet unless the orbital period is less than approximately 10 days \citep{2021A&A...652A.138S} or the Roche lobe filling factor is above $\sim0.8-0.9$ \citep{2021PASA...38...56H}.

OB stars in HeS binaries rotate fast (through Case B mass transfer) or slow (through Case A mass transfer), similar to OB+cc binaries. However, the colors of such binaries are generally bluer than OB+cc binaries due to the presence of the HeS components. OB+HeS binaries may be best detected by UV or X-rays observations \citep{2017A&A...608A..11G,2018A&A...615A..78G,2023ApJ...959..125G}.

\begin{figure*}[htbp]
\centering
\includegraphics[width=\linewidth]{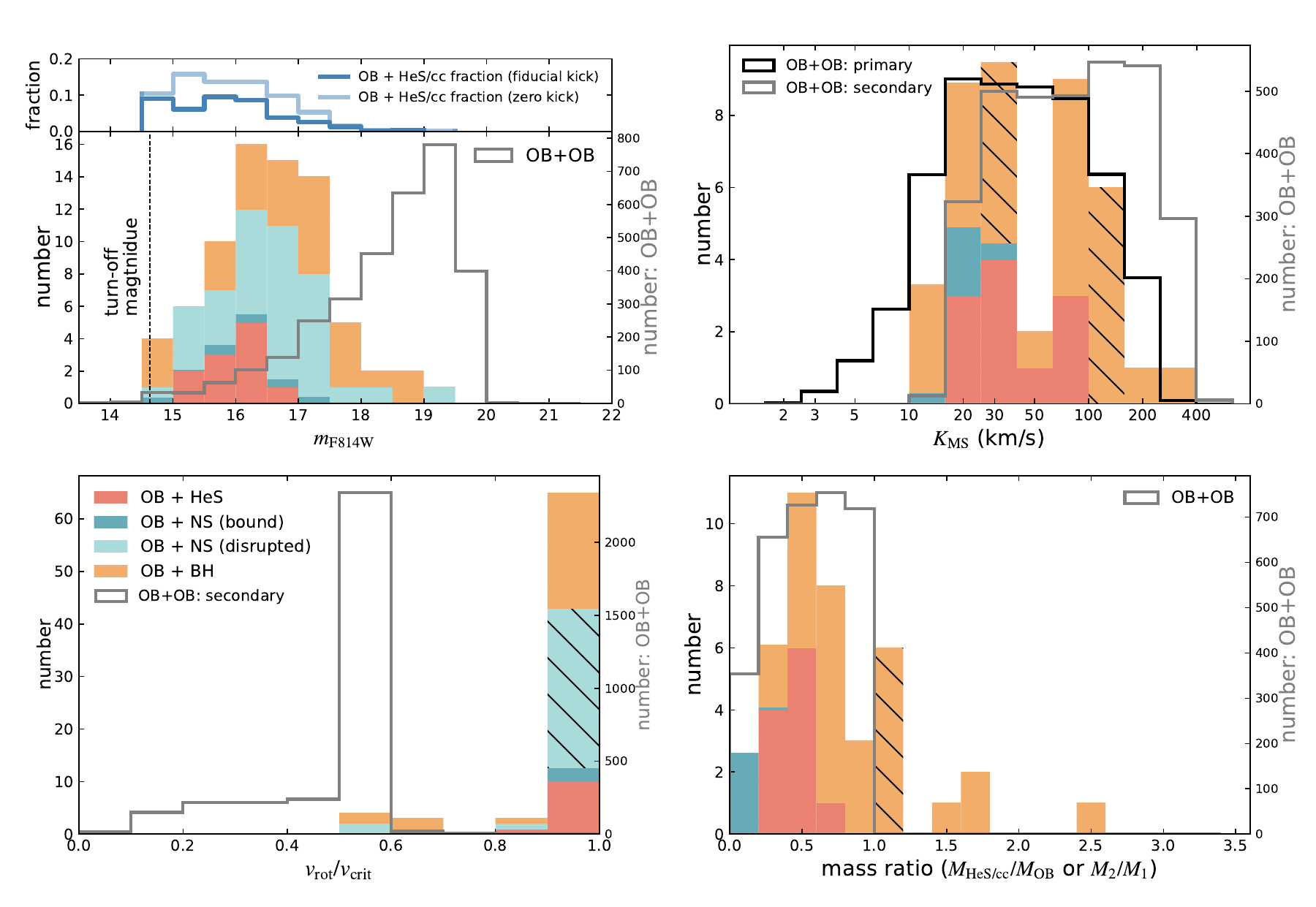}
\caption{Predicted properties of OB-type main-sequence star binaries containing a helium burning star (red-brown bars), a neutron star that remains bound after supernova kicks (turquoise bars with circles), a neutron star that is kicked out of the binary during supernova explosion (cyan bars with diagonal lines) or a black hole (orange bars) at 15\,Myr.
In the second and fourth panels, we exclude disrupted binaries to focus on binary features.
The first panel displays the number distribution as a function of magnitude in terms of the Hubble Space Telescope/WFC3 F814W band. The grey open bars denote binaries containing two OB-type main-sequence stars, with values on the right y-axis. The vertical dashed line marks the cluster turn-off magnitude. The upper panel shows the fraction of evolved massive binaries relative to the total number of stellar models at corresponding magnitudes, with (dark) and without (light) supernova kicks.
The second panel shows the semi-amplitude orbital velocity distribution of the OB star within evolved binaries and OB+OB systems.
The third panel illustrates the velocity distribution of OB star components in evolved massive binaries. Open grey bars denote the rotational velocities of the secondary stars in OB+OB binaries, shown on the right y-axis. The fourth panel presents the mass ratio distribution. For evolved massive binaries, the mass ratio is defined as the ratio of the post-main sequence component's mass to the OB component's mass. For OB+OB binaries, it is the ratio of the secondary star's mass to the primary star's mass.
}
\label{fig:BH_MS_15}
\end{figure*}


The upper left panel of Fig.\,\ref{fig:BH_MS_15} shows that the absolute number of these evolved binaries increases with magnitude up to around 16.5 magnitude, due to the effect of the IMF, and decreases at higher magnitude due to the fact that in our models, low-mass OB stars are more likely to merge during mass transfer. The upper small panel shows that the predicted fraction of evolved massive binaries is always smaller than approximately 20\% of all binaries at corresponding magnitudes, even when assuming no kicks in SN explosions. This is significantly lower than the observed Be star fraction of around 50\% near the turn-off region of young open clusters \citep{2018MNRAS.477.2640M}. Although Be stars near the cluster turn off can form from single star evolution, its contribution is largely confined to within approximately 0.5 magnitudes below the cluster turn-off \citep{2023A&A...670A..43W}. Consequently, the observed high frequency of fainter Be stars (fainter than 0.5 magnitudes below the cluster turn-off) suggests a need for refinement of mass transfer stability in our binary models (see Sec.\,\ref{sec:discussion_MT}).  

The lower left panel of Fig.\,\ref{fig:BH_MS_15} demonstrates that the majority of OB stars in evolved massive binaries in our simulations are rotating close to critical rotation. This agrees with the high fraction of Be stars observed in young open clusters \citep{Keller1999,Bastian2017,2018MNRAS.477.2640M}. Exceptions include those in Case A binaries and non-interacting binaries. 
For reference, we also present the distribution of rotational velocities for the secondary stars in pre-interaction OB+OB binaries. It can be seen that while most of the secondary stars in OB+OB binaries retain their initial rotation velocities, those in closer binary systems can undergo tidal spin-down. 

The upper right panel displays the distribution of the orbital velocity semi-amplitude. OB+BH binaries generally demonstrate higher values compared to OB+NS binaries, attributable to the higher masses of BHs. The range for semi-amplitude of orbital velocities in OB+BH binaries is quite extensive, varying from 10 to 300$\kms$, with two distinct peaks around 10 -- 40$\kms$ and 70 -- 120$\kms$, corresponding to binaries that have experienced Case B and Case A mass transfer, respectively. OB+NS binaries show orbital velocity semi-amplitudes ranging from 10 to 40$\kms$. The orbital velocity semi-amplitude for OB+He binaries falls within a range of 16 to 100$\kms$. Nowadays spectroscopic observations are able to identify OB binaries with semi-amplitude orbital velocities of approximately $10\kms$ or higher as single-lined spectroscopic binaries (SB1s), depending on the quality of the data, the spectral type of the star, and its projected rotational velocity \citep{2013A&A...550A.107S}. Consequently, nearly all evolved massive binaries predicted in our simulations at this age are expected to be identified as SB1s, given several epochs of observations covering the binary period.

The lower right panel of Fig.\,\ref{fig:BH_MS_15} illustrates the distribution of mass ratios ($M_{\rm HeS/cc}/M_{\rm OB}$) in these binary systems. OB+BH binaries exhibit a broad range of mass ratios (from $\sim 0.4$ to 2.5). High mass ratio OB+BH binaries evolve from binaries with initially large primary star masses but small secondary star masses (i.e. low initial mass ratios). 
For OB+NS binaries, mass ratios are generally below 0.2, as high mass ratio OB+NS binaries require low initial mass ratios, which can only undergo stable mass transfer with longer orbital periods (see Fig.\,\ref{app_fig:init_par}). However, such binaries are prone to disruption during SN explosions.
At 15\,Myr, all OB+HeS binaries are predicted to have mass ratios below 0.6, with progenitors having primary masses of around $14.5\mso$. At this mass, only binaries with initial mass ratios exceeding 0.55 can undergo stable mass transfer, leading to relatively massive OB stars in OB+HeS systems.

This section focused on a 15\,Myr-old coeval population. We display the CMD distribution and properties of our evolved massive binary models for further selected ages (from 10 to 100\,Myr) in Appendices\,\ref{app_sec:other_CMD} and \ref{app_sec:other_ages}, respectively.

\subsection{Incidence of evolved massive binaries as a function of cluster age}\label{sec:result_number}

\begin{figure*}[htbp]
\centering
\includegraphics[width=1.0\linewidth]{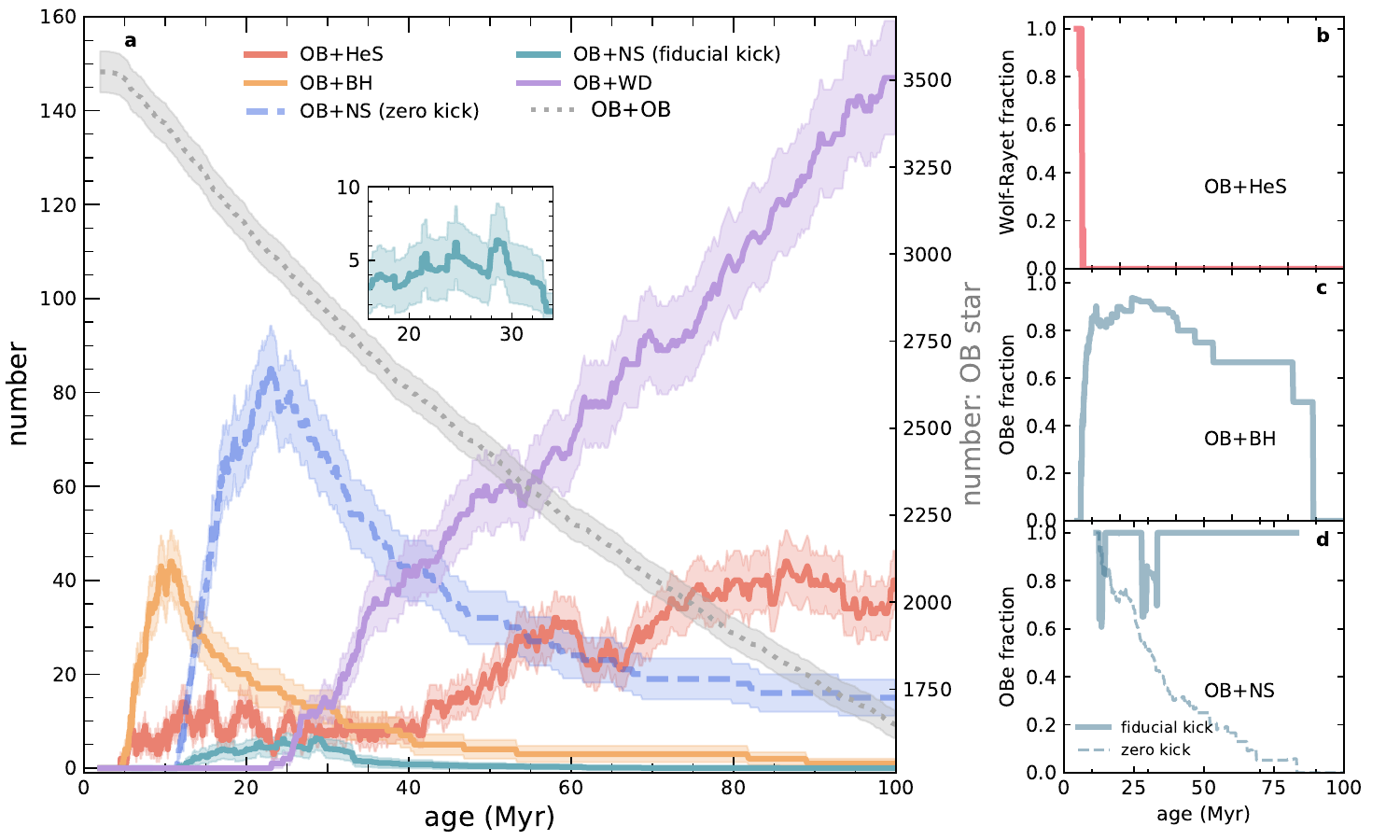}
\caption{Panel a: Predicted numbers of post-interaction binaries as a function of age, in a coeval stellar population with a total mass of $1.3\times 10^{5}\,$M$_{\odot}$. The solid lines in red brown, orange, turquoise and purple represent binaries with an OB star paired with a helium burning star, black hole, neutron star and white dwarf component, respectively, under the fiducial assumptions. The dashed blue line denotes the number of OB+neutron star binaries without including supernova kicks.
Additionally, we depict the number of OB+OB binaries  with $m_\mathrm{F814W}\leq 20$ with the dotted grey line, referring to the scale on the right y-axis for its values. The shaded areas show one standard deviation of the Poisson distribution. The inset provides a zoomed-in view of the results for OB+NS binaries, assuming fiducial kicks. Panel b: Fraction of Wolf-Rayet stars, which are identified according to the criterion in \cite{2020A&A...634A..79S},
among helium-burning stars with OB companions. Panel c: Fraction of OBe stars among OB+black hole binaries. Here, we treat stellar models whose surface rotational velocities are higher than 95\% of their break-up values as OBe stars. Panel d: Fraction of OBe stars among OB+neutron star binaries. The solid and dashed lines correspond to results for OB+neutron star binaries with and without supernova kicks.
}
\label{fig:number}
\end{figure*}

We create an animation showing the evolution of our models in the CMD over time, from 2 to 100\,Myr (Fig.\,\ref{fig:xxu_animation} is a snapshot at 15\,Myr). The time intervals in the animation are non-uniform to ensure smooth movement of the cluster turn-off and other binary evolutionary induced features, with higher resolution at younger ages.
The animation demonstrates the emergence and persistence of various types of evolved massive binaries in young open clusters. Figure.\,\ref{fig:number} plots the number of different post-interaction binaries over time, alongside the number of pre-interaction OB+OB binaries with $m_\mathrm{F814W}\leq 20$.

As expected, OB+HeS binaries are the first to appear. The time at which OB+HeS binaries begin to appear corresponds to the age of the most massive binary in which the primary star has completed its MS evolution and has avoided merging. In our simulations, two OB+HeS binaries appear around 4.5\,Myr. In one OB+HeS binary, the primary star has undergone a rotationally-induced chemically homogeneous evolution (CHE), i.e. the evolution of stars that are well-mixed, in which the stars have prolonged MS lifetimes and avoid a mass transfer process \citep{2009A&A...507L...1D,2016MNRAS.460.3545D,2016A&A...588A..50M}. The progenitor of this binary has $M_{\rm 1,i}=70\mso$, $M_{\rm 2, i}=17\mso$ and $P_{\rm i} = 38$\,d. The other one evolves from the binary with $M_{\rm 1,i}=51\mso$, $M_{\rm 2, i}=36\mso$ and $P_{\rm i} = 6$\,d through mass transfer.
The exact time is influenced by the initial mass of the most massive binaries, which can vary quite significantly due to the rarity of very massive stars as dictated by the IMF, coupled with small total number of massive stars in young open clusters.

Before $\approx$40\,Myr, the number of OB+HeS binaries remains below 20. However, after 40\,Myr, it increases to about 40 and stabilizes after $\approx$60\,Myr. 
Their number is influenced by several factors: the IMF and longer HeS lifetimes in less massive binaries tend to increase their numbers over time, while a lower chance of successful mass transfer and longer MS lifetimes before evolving into HeS phase in less massive binaries reduce them over time. These opposing factors shape the overall count of OB+HeS binaries, with fluctuations attributed to the limited sample size (cf., Sect.\ref{sec:discussion_size}).

Panel b of Fig.\,\ref{fig:number} shows the fraction of WR binaries among all HeS binaries as a function of cluster age. Here, we distinguish WR stars from normal HeS based on their luminosities. We classify HeS with $\log L \ge 5.6$ as WR stars, following \cite{2020A&A...634A..79S} for SMC stars. In our simulations, WR stars are only observed before approximately 7\,Myr. It is worth noting that we only consider WR stars with OB star companions. \cite{2003A&A...400...63V} argued that binary mergers can give rise to single WR stars as old as approximately 10\,Myr. The age of clusters or galaxies hosting WR stars is also influenced by metallicity. This is not only due to metallicity-dependent stellar winds, but also because of the varying luminosity thresholds for the WR phenomenon \citep{2020A&A...634A..79S}.

OB+cc binaries form when the primary star evolves into a compact object and persist until the secondary star completes its MS evolution. Binaries with smaller mass ratios, which successfully evolve into such systems, generally have longer lifetimes. However, these same binaries are also more susceptible to merging early on during the mass transfer process. 
In our simulations, the first OB+BH binary appears around 4.7\,Myr, originating from the above-mentioned OB+HeS binary that has undergone CHE. The number of OB+BH binaries peaks around 11\,Myr, with a value of over 40. Prior to this peak, there is an increase in numbers, due to the growing number of progenitors, an effect of the IMF and increasing stellar lifetime. Beyond the peak, there is a decline in numbers, as the nature of the newly formed compact objects shifts. Specifically, after 11\,Myr, the majority of new compact objects tend to be NSs rather than BHs. However, the previously formed OB+BH binaries remain in this category until their OB companions end their MS evolution. One BH is predicted at 100\,Myr, it evolves from a non-interaction binary with an initial orbital period of 2118\,d. This system may resemble the long-period Gaia BH3 with high eccentricity, except that its companion is more massive. The oldest cluster hosting OB+BH binaries that have experienced a stable mass transfer has an age of $\approx$90\,Myr. Our oldest post-interaction BH-binary has initial parameters of $M_{\rm 1,i}=20\mso$, $M_{\rm 2, i}=5$, and $P_{\rm i} = 337$\,d and current parameters of $M_{\rm BH}=5.5\mso$, $M_{\rm 2}=5$, and $P = 12$\,d.

In Panel c of Fig.\,\ref{fig:number}, we display the fraction of OBe stars among OB+BH binaries. Here, we assume that stars rotating faster than 95\% of their break-up velocities are OBe stars. It is important to note that Be stars may rotate subcritically, with rotational velocities as low as approximately 75\% of their critical rotational velocities \citep{2013A&ARv..21...69R}. We use a high threshold here to select stars that originate from binary evolution, as the majority of accretors in our binary models reach critical rotation after mass transfer. Using a lower threshold would slightly increase the number of OBe stars by including those come from single star evolution. However, as mentioned in \cite{2023A&A...670A..43W}, single star evolution may only contribute to the formation of Be stars within 0.5 magnitude below the cluster turn off. 
Panel c of Fig.\,\ref{fig:number} shows that the OBe fraction is very high before approximately 90\,Myr. These binaries are predicted to be X-ray quite due to their wide orbits \citep{2021A&A...652A.138S,2024arXiv240608596S}, but could be able to be detected as SB1s (see Fig.\,\ref{fig:BH_MS_15}.)

The first OB+NS binary in our simulation forms around approximately 11.5\,Myr, originating from a Case B binary with initial parameters of $M_{\rm 1,i}=18.7\mso$, $M_{\rm 2, i}=16.8$, and $P_{\rm i} = 11$\,d. From this point onward, the newly formed compact objects are predominantly NSs rather than BHs. 
For the number of OB+NS binaries, we consider both zero SN kicks for starbursts and fiducial kicks for star clusters. In the case of zero SN kicks, the predicted number of OB+NS binaries peaks around 22\,Myr. 
When considering SN kicks, the predicted number of OB+NS binaries is found to be less than six at all ages. The number distribution of these binaries exhibits several distinct small peaks, at approximately 17, 22, 24.5 and 28.5\,Myr, respectively (see the inset of Fig.\,\ref{fig:number}). The first two peaks align with the highest predicted number of OB+NS binaries prior to the inclusion of SN kicks. The subsequent two peaks are linked to stars that experience ECSN. In our assumption, ECSN events result in smaller SN kicks, thereby enhancing the likelihood of binary survival. There are only seven binaries undergo ECSN in our simulations.

In Panel d, we show the fraction of OBe stars among OB+NS binaries. In the case of zero kicks, the fraction of Be stars decreases with age because, at older ages, the contribution from non-interaction binaries increases as the parameter space for non-interacting binaries increases with decreasing masses. Conversely, in most cases with fiducial SN kicks, the predicted OBe star fraction is close to 1, as almost all long-period non-interaction binaries are destroyed during SN kicks. Between approximately 13 to 33\,Myr, Case A binaries can exist and remain bound after SN kicks, producing binaries containing a sub-critically rotating OB star. Therefore, the Be star fraction is smaller than 1 during this period.

The first OB+WD binary appears at roughly 23\,Myr, evolving from a Case A binary with $M_{\rm 1,i}=12.9\mso$, $M_{\rm 2, i}=8.1$, and $P_{\rm i} = 1.5$\,d. Following this time, 
the newly formed compact objects are WDs rather than NSs. 
The number of OB+WD binaries show a linear increase over time, due to the fact that from this age forward, newly formed compact objects in binaries mainly evolve into WDs.

The predicted number of evolved massive binaries can be significantly influenced by uncertainties regarding mass transfer stability, SN explosions and sample size. We discuss these uncertainties in Sec.\,\ref{sec:discussion} and Appendices\,\ref{app_sec:MT}, \ref{app_sec:SN} and \,\ref{app_sec:size}.

\subsection{Properties of evolved massive binaries as a function of cluster age}
\begin{figure*}[htbp]
\centering
\includegraphics[width=\linewidth]{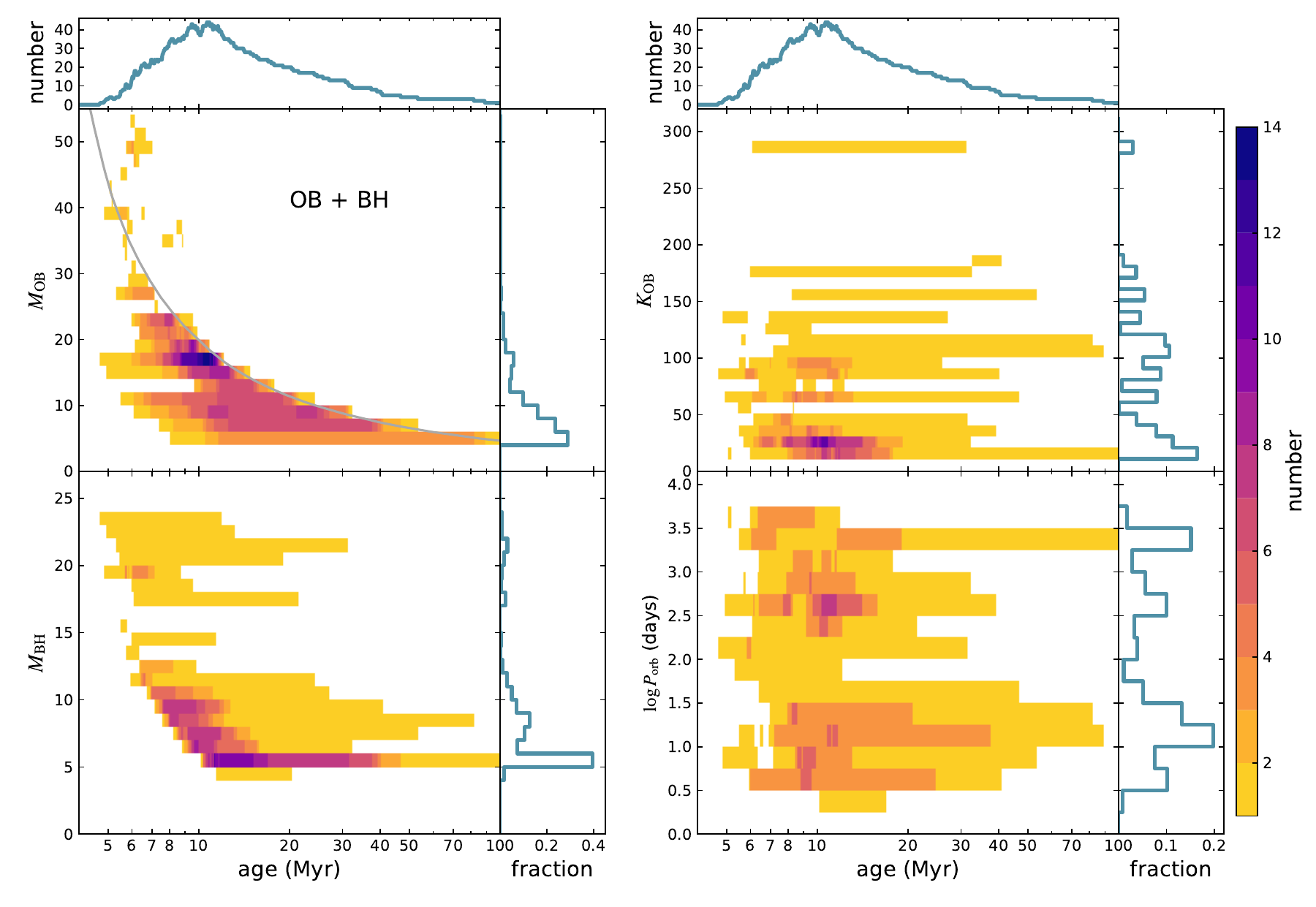}
\caption{Predicted properties of OB+BH binaries as a function of cluster age. The color coding represents the number density of these binaries in each pixel.
The left panels display the mass distribution of OB stars and BHs, whereas the right panels show the distribution of the semi-amplitude orbital velocity of OB stars and current binary orbital periods. The grey line in the upper left panel depicts the turn-off mass as a function of cluster age.
The step plots at the top and right margins of the main panels represent 1D projections. Notably, the 1D projection on the right side is normalized to unity, taking into account the duration of binaries possessing specific properties.
}
\label{fig:BH_MS}
\end{figure*}

In Fig.\,\ref{fig:BH_MS}, we use OB+BH binaries as examples to display their properties as a function of age, up to 100\,Myr, as predicted by our models. This includes the distributions of OB star masses, BH masses, orbital periods and the semi-amplitude of the orbital velocities. 

The upper left panel reveals that the masses of OB stars in these binaries are generally constrained by the cluster turn-off mass, indicated by the grey line. This is because most OB+BH binaries originate from highly non-conservative Case B mass transfer, meaning that the OB stars experience little to no rejuvenation. The exceptions are mainly binaries that have experienced Case A mass transfer, which can significantly rejuvenate the OB stars and extend their MS lifetimes. In our simulations, there is also one massive binary, in which both stellar components experience CHE and rejuvenation. In this binary, the OB star surpasses the cluster turn-off mass during OB+BH phase. However, in most cases, CHE only happens in the massive primary star and the OB companion evolves normally without experiencing rejuvenation.

There is a trend in the OB star mass distribution: for a given OB star mass, the number of OB+BH binaries increases over time (i.e. as one approaches the grey line). 
This occurs because when an OB star enters the OB+BH phase at a certain age, it maintains its mass until the end of its MS life due to weak stellar winds in SMC stars. During this period, additional OB stars of similar mass may also evolve into the OB+BH phase, thereby enhancing the number of OB stars. This trend is also an effect of the IMF. Over time, for a given OB star mass, binaries with lower initial primary masses, favored by the IMF, will evolve into OB+BH binaries with that specific OB star mass.  

The lower left panel demonstrates that the BH masses in OB+BH binaries predominantly range from 5 to $10\mso$. Unlike the masses of OB stars, the mass of BHs can exceed the cluster's turn-off mass. For instance, a 20\,Myr-old cluster is capable of hosting binaries with BHs exceeding 20$\mso$. In a star cluster, the maximum mass BH is produced by a binary in which the secondary OB star's mass equals the cluster turn-off mass, while the primary star's mass is the maximum allowed for stable mass transfer.
This panel shows that there is an overall trend that older clusters tend to have less massive BHs. This is because the OB stars in older clusters have lower masses, and for stable mass transfer to occur, their BH companions should also be less massive. In contrast to the trend seen with OB stars, the number of OB+BH binaries for a given BH mass tends to decrease over time. The reason is that as existing binaries evolve beyond their OB+BH phases, no additional binaries with similar BH masses enter this category.

The present-day distributions of orbital period and semi-amplitude for OB+BH binaries predicted by our models display a bimodal pattern, due to two distinct mass transfer scenarios (Case A and Case B mass transfer). The semi-amplitude orbital velocity distribution for OB stars shows peaks around $10-50\kms$ and $60-120\kms$, corresponding to binaries formed through Case B mass transfer (or non-interaction binaries) and Case A mass transfer, respectively. These findings agree well with the results in \cite{2020A&A...638A..39L,2024Xu}. Given the current threshold of approximately 10$\kms$ for systems to be identified as SB1s using spectroscopic observations \citep{2013A&A...550A.107S}, we conclude that nearly all OB+BH binaries predicted in our simulations are potential candidates for detection. However, this depends on the quality of the data, the spectral type of the star, and its projected rotational velocity.

We display the predicted properties of OB+HeS, OB+NS and OB+WD binaries in Appendix\,\ref{app_sec:other_binaries}. In particular, we find that the mass of HeS in OB+HeS binaries follows the cluster turn-off mass well, indicating that OB+HeS binaries could be used as an independent measure for cluster age.

\section{Comparison with observations}\label{sec:compare_ob}
The comparison between detailed binary models and field stars, including WR stars and Be X-ray binaries, is explored by \cite{2024Xu}. In this section, we focus on comparing our theoretical predictions with observations in young open clusters. 
Despite the challenges of performing multi-epoch spectroscopic observations for individual stars in young open clusters, significant progress has been made.

The SMC cluster NGC\,330 is of particular interest as its metallicity matches our binary models and its age (approximately 30 to 40\,Myr) is covered by our simulations. \cite{2020A&A...634A..51B,2021A&A...652A..70B,2023A&A...680A..32B} measured stellar properties for individual OB-type stars in this cluster with six epochs of MUSE spectroscopic observations. 

These studies identify 115 Be stars at the upper right side of the cluster turn-off. The positions of these Be stars in the CMD align well with both photometric observations and our model predictions \citep{2018MNRAS.477.2640M,2022NatAs...6..480W}. However, both the observed absolute number and fraction of Be stars are higher than our predictions. According to \cite{2020A&A...634A..51B}, NGC\,330 contains 412 stars with masses between 3 and 5$\mso$ in NGC\,330. In our simulations, at the age of NGC\,330, we predict 1855 pre-interaction OB+OB binaries whose primary masses are within this range. We predict 71 Be stars (assuming stars rotating faster than 75\% of their critical velocities as Be stars), with 70 of them resulting from binary evolution and 35 possessing NS companions. After adjusting our predictions to match the observed number of stars in the 3 to 5$\mso$ range, we estimate only around 16 Be stars, half of which may be disrupted by SN kicks and ejected from the cluster. This number is significantly lower than the observed 115 Be stars. Reducing the rotational velocity threshold from 75\% to 70\% of the critical velocity only slightly increases the predicted number of Be stars (from 71 to 81). However, under the idealized assumption that all binaries undergo stable mass transfer and each produces one Be star, the predicted number of Be stars aligns well with the observed number (see Fig.\,\ref{app_fig:MT_test}).

The detected fraction of Be stars (32\% for stars with $m_\mathrm{mF814W}\leq 19$, and $\sim 50\%\pm 10\%$ near the cluster turn off) is also significantly higher than predicted by our models (less than approximately 10\%). These discrepancies indicate the need to enhance the stability of mass transfer in our models, thereby increasing the contribution from binary evolution. Additionally, the role of single stars requires further investigation. Although our simulations show that single star evolution does not significantly contribute at this age, the outcomes are highly sensitive to factors such as the initial rotational velocities of stars, the stellar wind mass-loss rates, and the velocity threshold for a star to become a Be star \citep{2023A&A...670A..43W}. Moreover, other mechanisms, such as stellar pulsations, may also play a role in Be star formation \citep{2013A&ARv..21...69R}. Given that the Be star phenomenon and its origins are not yet fully understood, further studies are urgently needed.

\cite{2023A&A...680A..32B} found that the mean projected rotational velocity of Be stars is significantly larger than that of normal B-type stars, which agrees with our understanding of Be stars. Among 115 Be stars, 93 have radial velocity measurements, and 7 are detected as binaries, with an overall spectroscopic binary fraction of $7.5\pm2.7\%$ \citep{2021A&A...652A..70B}. 
Our simulation predicts that at the age of NGC\,330, 53\% of the Be stars have NS companions without including SN kicks. Approximately 90\% of these binaries are broken up by SN kicks, leaving single Be stars. Systems that remain bound are potential progenitors of X-ray binaries. Indeed, an X-ray source was detected in the core of NGC 330 \citep{2005MNRAS.362..879S,2013A&A...558A...3S,2016A&A...586A..81H}, and, based on the MUSE data, \cite{2020A&A...634A..51B} reported that the two stars near this X-ray source are both Be stars. While one star shows no significant RV variations, the other one exhibits RV variations measured from the strong emission lines (those impact also the He I lines commonly used for RV measurements). Unfortunately, current observations cannot unambiguously link these stars to the X-ray binary source. Follow-up observations are necessary.

We predict about 19\% of Be stars to have WD companions. However, due to the low mass of WDs, the predicted semi-amplitude orbital velocity is always smaller than 15$\kms$, which is close to the lower observational boundary of 10$\kms$ for identifying SB1s. Meanwhile, we also predict that 19\% of Be stars have BH companions, with their semi-amplitude orbital velocities exceeding the observational threshold, indicating that these could be detectable as SB1s. Although some Be stars may originate from single star evolution, this only occurs near the cluster turn-off \citep{2020A&A...633A.165H}. Be star below the cluster turn-off can only be explained by binary evolution (see Fig.\,\ref{fig:xxu_animation}). Therefore, we predict at least 19\% of Be stars to be detectable as SB1s. The observed fraction of $7.5\pm2.7\%$ aligns reasonably well with our predictions, especially when considering that this fraction has not been corrected for observational biases.

In addition to the SMC cluster NGC\,330, \cite{2023MNRAS.518.1505K} have conducted 16-epoch observations of stars in the $\approx 100$\,Myr-old cluster NGC\,1850 in the LMC.
With these observations, \citet{2023MNRAS.518.1505K} identified more than 200 fast-rotating Be stars. A spectroscopic binary fraction of $10.0\pm 2.8\%$ has been measured \citep{2023MNRAS.526..299S} among these Be stars, which is comparable to the fraction measured in NGC\,330 \citep{2021A&A...652A..70B}. Interestingly, NGC\,1850 BH1 was initially claimed to be a binary containing a BH of around 11.1$\mso$ \citep{2022MNRAS.511.2914S}, but later argued to be a stripped star of approximately 1$\mso$ masquerading as a BH \citep{2022MNRAS.511L..24E}. This binary system appears similar to the binaries LB-1 and HR\,6819, except the MS companion is not a Be star. 
Our simulations indeed predict several binaries that can roughly match the observed properties of NGC\,1850 BH1. Detailed comparisons are left for future studies.
Although such detections are still rare, they are crucial for advancing our understanding of binary evolution.

In addition to these two clusters with available multi-epoch data, multi-epoch observations of stars in more clusters, covering a wide range of ages 
have been achieved or approved. 
Studies of these clusters will soon provide us numerous data to compare with our models, thereby enhancing our understanding of massive binary evolution.

\section{Discussions}\label{sec:discussion}
\subsection{Effect of limited sample size}\label{sec:discussion_size}
In this study, we utilize a collection of 3670 detailed binary-evolution models, whose initial parameters are created using a Monte Carlo method taking into account empirical binary parameter distributions. Although the sample size may seem limited, it is exceptionally well-suited for this research. Our simulations represent a star cluster with a total stellar mass of $1.3\times 10^5\mso$, which corresponds to the typical upper mass limit for young open clusters in the Magellanic Clouds \citep{2003AJ....126.1836H,2005ApJS..161..304M,2012ApJ...751..122P}. Moreover, the chosen sample size guarantees optimal binary representation across our considered parameter space. This can be seen by the smooth evolution of our models in the animation, as well as the parameter distribution in Fig.\,\ref{app_fig:init_par}.

The models used in this work are based on our previously computed large grid of detailed binary evolution models \citep{2020ApJ...888L..12W} with constant initial parameter intervals. The effect of binary evolution in producing multiple MS components in young open clusters observed in this work aligns well with \cite{2020ApJ...888L..12W}. The advantage of the models used in this work over the large grid in \cite{2020ApJ...888L..12W} is that they allow us to eliminate the need for intricate interpolations among binary models in population synthesis simulations. This is particularly important when examining binaries near the cluster turn-off, where outcomes are sensitive to binary parameters. With these models, we are able to create an animation in which the distribution of our models in the CMD evolve smoothly with time.

To explore the effect of limited sample size in our study of young open clusters, we perform additional analysis in Appendix\,\ref{app_sec:size}, where we reduce the sample size to represent less massive clusters. We find consistent patterns in the predicted numbers of evolved massive binaries until the sample size is reduced to one-fifth of the original. This uniformity underscores the effectiveness of our binary sampling. 

As for the study of starbursts, it is the evolutionary pattern of the number of evolved massive binaries and their properties, rather than the absolute number, that matters. Therefore, the limited sample size will not impact our conclusions for the studies of galaxies with starbursts.

\subsection{Effect of initial binary parameters}
The initial binary parameter distributions have significant influence on the predicted stellar populations in young star clusters. In the context of star clusters, the primary star mass is the key parameter in determining if a binary system has engaged in mass transfer or not. Binaries with initial primary masses notably below the cluster turn-off mass will most likely remain as pre-interaction binaries, unless they have exceptionally short orbital periods. Conversely, binaries with initial primary masses exceeding the cluster turn-off mass are typically post-interaction systems, with some potentially being semi-detached binaries.

In our simulation, we adopt the Salpeter IMF with a power-law exponent of -2.35 for the primary mass distribution. This choice is supported by observations of OB associations and clusters in the Local Group, which have revealed similar power-law exponents as the Salpeter IMF \citep{2003ARA&A..41...15M,2004ApJ...617.1115D,2000A&A...353..655K}. Meanwhile, in \citet{2022NatAs...6..480W}, we derived similar exponents for red MS stars in young open clusters. However, it is worth noting that a slightly top-heavy IMF with a power-law exponent of $-1.9^{+0.37}_{-0.26}$ was found for O-type stars in the LMC 30 Doradus region \citep{2018Sci...359...69S}. Adopting such a top-heavy IMF would result in a larger number of evolved massive binaries in younger clusters.

Mass ratio and orbital period are crucial in determining stable or unstable mass transfer. In our models, higher mass ratios and larger orbital periods favor stable mass transfer. We use a flat distribution for both mass ratio and orbital period in logarithmic space. However 
studies like \citet{2012Sci...337..444S} suggest power-law distributions for both mass ratio ($f(q)\propto q^{\kappa}$ with $\kappa=-0.1\pm 0.6$ across $0.1<q<1$) and orbital period ($f(\log P)\propto \log P^{\pi}$ with $\pi = -0.55$ for $1.4 < P /\mathrm{days} <1000$) based on spectroscopic measurements of Galactic massive O-type stars. Their mass ratio distribution is similar to the one we use, but their orbital period distribution skews towards short period. Incorporating this would likely lead to an increased number of Case A mass transfer. This would result in an overall decrease in OB+BH binaries and increase in OB+NS binaries, due to the fact that the helium core mass of the primary star is smaller after Case A mass transfer than Case B mass transfer. In addition, more Case A binaries would give rise to more shorter-period binaries with more massive OB stars due to effective rejuvenation. Meanwhile, a reduction in long-period Case B binaries would lead to a decrease in the number of less massive OB stars in evolved massive binaries at a given age, since for a fixed mass ratio, binaries with longer periods are more likely to undergo stable mass transfer.

\citet{2017ApJS..230...15M} found a mass-dependent orbital period distribution, where stars more massive than 5$\mso$ tend to have shorter periods, whereas stars less massive than this show a preference for longer periods \citep[see also][]{2024arXiv240601420S}. This can impact the rate of double BH and NS mergers \citep{2018MNRAS.481.1908K,2018A&A...619A..77K,2021ApJ...922..110G}. Their relation suggests more Case A mass transfer, which leads to an increase in shorter-period binaries with more massive OB stars in younger clusters, and more Case B mass transfer, which results in longer-period binaries with less massive OB stars in older clusters.


\subsection{Effect of mass transfer stability and efficiency}\label{sec:discussion_MT}
Mass transfer stability is one of the major uncertainties in binary evolution. 
When computing our binary models, we incorporate a more physically motivated prescription that considers the spin-up of accretors, instead of using a simple criterion based on mass ratio for deciding mass transfer stability, as is common in the literature. With this approach, binaries with higher primary masses, higher mass ratios and longer orbital periods are prone to experience stable mass transfer \citep{2020A&A...638A..39L,2024Xu}. Those that undergo unstable mass transfer will merge and produce blue stragglers \citep{1964MNRAS.128..147M,1999ApJ...513..428S,2001ApJ...552..664N,2016MNRAS.457.2355S,2022NatAs...6..480W} or blue supergiants in young open clusters \citep{1990A&A...227L...9P,2014ApJ...796..121J,2024ApJ...963L..42M}.

However, in order to explain the observed high number of Be X-ray binaries in the SMC \citep{2016A&A...586A..81H} and the high fraction of Be stars in young open clusters in the Magellanic Clouds \citep{2018MNRAS.477.2640M}, a larger fraction of binaries than predicted by our models must undergo stable mass transfer \citep{2021A&A...653A.144H,2022NatAs...6..480W,2024Xu}. There are various ways of achieving this: 
\begin{itemize}
\item Mechanisms that allow for continuous accretion even after an accretor achieves critical rotation, including transport angular momentum outwards through accretion disks \citep{1991ApJ...370..604P,1991ApJ...371L..63P}.
\item Unique energy sources capable of expelling excess material from stars, like the radiation emitted from the hot spot where the accretion stream collides with the mass gainer \citep{2011A&A...528A..16V}.
\item The radiation energy available for expelling material may be underestimated in our models. Instead of leading to an immediate merger, the accreted material might remain around the system until the primary star re-establishes equilibrium and emits a substantially enhanced luminosity \citep{Pauli2020}.
\item Magnetic fields may be generated on the surface of the accretor and prevent or delay the accretor from reaching critical rotational velocity \citep{2010MNRAS.406.1071D}.  
\end{itemize}

Considering our strict criteria for binary mergers, the predicted number of OB+HeS/cc binaries in Fig.\,\ref{fig:number} should be viewed as a lower limit. In Appendix\,\ref{app_sec:MT}, we examine an idealized case, in which all Case B binaries are assumed to undergo stable mass transfer and lead to an OB+cc binary \citep{2021A&A...653A.144H}. Figure\,\ref{app_fig:MT_test} displays the newly predicted numbers for OB+cc binaries. It can be seen that the distribution trend and peak positions of OB+cc binaries remain unchanged, but the predicted number increases, especially dramatically in OB+WD binaries. Our results imply that identifying various types of OB+cc binaries in young open clusters of different ages may provide useful constraints on mass transfer stability. 

Mass transfer efficiency remains one of the least constrained parameters in binary evolution. While some observations support non-conservative mass transfer, others can be explained by near-conservative mass transfer \citep[see][and references therein]{2012ARA&A..50..107L}. Specifically, recent discoveries of OB binaries with hot stripped stars suggest a need for higher mass transfer efficiency than currently assumed in our simulations \citep{2023AJ....165..203W}. In our simulation, the efficiency of mass accretion is closely tied to the binary orbital periods. For Case B binaries, mass transfer is highly non-conservative, with efficiencies lower than 10\%, while Case A binaries can achieve efficiencies up to 60\% \citep{2022A&A...659A..98S,2023A&A...672A.198S}. This correlation between mass transfer efficiency and orbital periods agrees with the findings in \citet{2007A&A...467.1181D}. 
Mass transfer efficiency has a large impact on the predicted mass of OB stars in evolved massive binaries, with higher mass transfer efficiency resulting in more massive OB stars. This may also affect the predicted number of evolved massive binaries in young open clusters as more massive OB stars have shorter MS lifetimes. Nevertheless, the influence of the uncertainties of mass transfer efficiency on the predicted number is much less significant than the uncertainties of mass transfer stability.

\subsection{Effect of assumptions regarding compact object formation}\label{sec:discussion_SN}
In our fiducial simulation, aligned with the assumption in our series of papers \citep{2020A&A...638A..39L,2024Xu}, we adopt a simplified approach for compact object formation. This approach does not account for the non-monotonic pattern of NS and BH formation \citep[see e.g. ][]{1996ApJ...457..834T,2011ApJ...730...70O,2014ApJ...783...10S,2016ApJ...821...38S,2018ApJ...860...93S,2016ApJ...818..124E,2019ApJ...878...49W,2020ApJ...890...43C,2020A&A...637A...6L}. This approach also ignores the argument that the core structures of binary-stripped stars are systematically different to those of single stars \citep{2021A&A...656A..58L,2021A&A...645A...5S}. \citet{2021A&A...645A...5S} investigated the end states of binary stripped stars and found that they can evolve to BHs only if the CO core mass exceeds 15$\mso$ or lies between 6 and 8$\mso$ at helium exhaustion. Otherwise, a successful SN event is expected to happen and produce a NS. We examine the implications of this refined understanding of compact object formation on our findings in Appendix\,\ref{app_sec:SN}.

Figure\,\ref{app_fig:SN_Fabian} shows that adopting the criteria from \citet{2021A&A...645A...5S} leads to a decrease in the number of predicted OB+BH binaries. This is because many primary star models within the initial mass range of $30\mso$ to $65\mso$ are now more likely to result in NSs rather than BHs. But the predicted BH masses are larger than in our fiducial results. In addition, OB+NS binaries are predicted to emerge earlier, starting from approximately 5\,Myr. This experiment suggests that detecting compact objects in clusters with ages between 5 to 10\,Myr may enhance our understanding of compact object formation. In particular, the discovery of the X-ray pulsar CXO J164710.2-455216 in the $\approx$5\,Myr-old cluster Westerlund 1 \citep{2006ApJ...636L..41M} provides evidence that stars more massive than $\approx 40\mso$ can undergo successful SN explosions and form NSs.

Furthermore, using stripped-star models, \cite{2023A&A...671A.134A} found that after considering fallback according to \cite{2020MNRAS.499.3214M}, several models with CO core masses up to approximately $30\,\mso$ may form low-mass BHs (approximately $3\,\mso$) through successful explosion and fallback. These models would otherwise produce high-mass BHs in our fiducial models and models in \cite{2021A&A...645A...5S}. Although this fallback effect does not change the number of OB+BH binaries, it influences their properties in very young clusters by producing low-mass BHs. Additionally, \cite{2023A&A...671A.134A} also found that the explodability and the properties of compact remnants are affected by metallicity due to metallicity-dependent stellar winds. They reported a weak trend indicating that, at high-final masses (larger than approximately $13\,\mso$), explodability increases with decreasing metallicity, favouring the formation of NSs. 
However, the lowest metallicity in their study is 0.01, lower than the value of 0.0142 used in \cite{2021A&A...645A...5S}, but still significantly higher than our binary models. Therefore, further investigation into the final evolution of stripped stars and their remnants at low metallicties are extremely important.

\subsection{Effect of triples and higher-order systems}\label{sec:discussion_triple}
The overall fraction of triple and higher-order systems amongst OB-type stars is found to be of the order of 50\% \citep{2017ApJS..230...15M,2023ASPC..534..275O}, which includes systems with separation of up to 10$^4$\,AU. Considering a rough separation limit of 100\,AU beyond which a dynamical interaction of the outer
companion in a triple system with the inner binary appears less likely would imply that this might happen in about two thirds of the systems. While this is still a high fraction, these measurements do not relate to the young, rich clusters which are investigated in this paper, but mostly to field stars. In particular, wide multiple systems might be disrupted in the cluster situation. Indeed \cite{2022NatAs...6..480W} analyzed high precision HST photometry data of four rich young LMC/SMC clusters \citep{2018MNRAS.477.2640M} and found the fraction of B star triple systems in all four cases to be smaller than 10\%.

The long-term evolution of these complex systems remains poorly understood \cite[see][for a review]{2016ComAC...3....6T}. The presence of a tertiary companion can alter the evolutionary path of the inner binary in two primary ways. Firstly, Roche-lobe overflow can initiate from the tertiary to the inner binary \citep{2020A&A...640A..16T,2022MNRAS.517.2111P,2024MNRAS.527.9782D}. However, this scenario only occurs under restricted conditions, where the outer orbit is sufficiently small for the tertiary to fill its Roche lobe. \cite{2020A&A...640A..16T} found that in systems with intermediate-mass primaries (1-7.5$\mso$), less than 1\% of triples experience mass transfer from the tertiary. For massive stars between 10 and 100$\mso$, \cite{2023A&A...678A..60K} found a slightly higher fraction of $1.8\pm 0.2\%$, which still remains a minor contribution to the overall population. 

In addition, the tertiary can impact the inner binary through dynamical interactions. One possible dynamical interaction is von Zeipel-Lidov-Kozai (ZLK) cycles \citep{1910AN....183..345V,1962P&SS....9..719L,1962AJ.....67..591K} that can induce oscillations of the inner binary. Such oscillations can lead to high eccentricities of the inner binary, which subsequently result in tidal interactions that shrink the inner binary's orbit. This process can enhance the rates of mass transfer and mergers in the inner binary. 

The evolution of inner binaries following ZLK cycles and tidal synchronization, as well as mergers, are already well-represented in our models. Their distributions in the CMD are captured effectively in our animation \footnote{However, it is worth noting that mass transfer in high eccentric binaries \citep{2007ApJ...667.1170S,2009ApJ...702.1387S,2016ApJ...825...70D,2016ApJ...825...71D,2020PASA...37...38V,2021MNRAS.507.2659G}, a phenomenon not included in our models, remains poorly understood and is beyond the scope of this work.}. Nevertheless, the overall number of predicted massive binaries may vary as a result of triple star evolution. 
For example, \cite{2020A&A...640A..16T} found that for intermediate-mass primaries (1-7.5$\mso$), ZLK cycles can increase the frequency of Case A mass transfer binaries by a factor of 3.7-5.5. This suggests that triple star evolution could lead to an enhanced population of blue stragglers in young open clusters \citep{2009ApJ...697.1048P,2022MNRAS.515L..50V}, as well as more evolved massive binaries with sub-critically rotating OB stars than predicted by our models. However, \cite{2020A&A...640A..16T} also found that the frequency of Case B mass transfer binaries remains largely unaffected. This is because mass transfer triggered in wider inner binaries compensates for the close binaries undergoing Case A mass transfer in triples. Since the majority of evolved massive binaries in our simulations originate from Case B mass transfer, we conclude that triple star evolution will not significantly alter our results and conclusions.

While the evolution of triple stars is undoubtedly important, fully accounting for it in population synthesis studies with detailed models remains extremely challenging. This is due to the vast parameter space involved as well as from our incomplete understanding of triple star evolution. As such, a more comprehensive investigation of triple star evolution is beyond the scope of this work and is left for future studies. 
\section{Concluding remarks}\label{sec:conclusion}
In this \textit{Letter}, we have provided predictions for the number and properties of stripped helium and compact object binaries with an OB companion in coeval stellar populations with ages up to 100\,Myr. The predictions are based on 3,670 detailed binary evolution models computed by the 1D stellar evolution code MESA. To compute these models, we consider the spin-up of accretors during the accretion process and determine binary mass transfer stability and efficiency in a self-consistent way. 
We take into account empirical initial binary parameter distributions. Our simulations represent clusters of total initial stellar masses of approximately $10^5 \mso$, and are easily scaled to masses of coeval populations in starbursts. 

As a case study, we have shown the CMD distribution and properties of our post-mass transfer binaries in a cluster of 15\,Myr (the estimated age for the LMC cluster NGC\,2100). Our results reveal that these binaries are located within 3-4 magnitudes below the cluster turn off.
In addition, the OB components in the majority of these binaries are near-critically rotating.

We have created an animation that shows the time evolution of our binary models in the CMD, which may serve as a roadmap for future observational efforts. 
Our simulations predict a notable number of OB+HeS/cc binaries in young stellar populations. Specifically, OB+BH binaries peak in number at approximately 10\,Myr, and post-interaction OB+BH binaries can persist in clusters/starbursts up to 90\,Myr (Fig.\,\ref{fig:number}). While we anticipate a significant number of OB+NS binaries, most are likely to be disrupted by SN kicks. Clusters/starbursts with ages between 15 to 35\,Myr are prime candidates for their detection, where the low kicks of ECSNe contribute notably to the surviving systems at ages near 27\,Myr. The emergence of OB+WD binaries starts around 23\,Myr, with their numbers increasing linearly with age. We also predict the existence of numerous OB+HeS binaries in young open clusters and starbursts. Specifically, we forecast a high fraction of WR stars among these binaries for populations younger than approximately 7\,Myr. However, this age limit is sensitive to the dependence of WR phenomenon on metallicity.

We have also deduced the distribution of masses and orbital velocities for these binaries. We found that, in general, the OB star mass in these binaries cannot exceed the cluster turn-off mass. The exceptions are those from chemically-homogeneous evolution and from Case A mass transfer. 
The mass of BHs in these binaries shows no correlation with the cluster turn-off mass. For example, a 30\,Myr-old cluster (such as SMC NGC\,330) can host BHs up to $20\mso$. Despite the likelihood of being X-ray quiet, the predicted semi-amplitude orbital velocities for OB+BH binaries indicate their potential detectability via spectroscopic methods.


Detecting evolved massive binaries is crucial for enhancing our understanding of massive star evolution and the progenitors of GWs. Our research provides a guide for observers, assisting them in refining their survey strategies for these targets. Multi-epoch spectroscopic observations for individual stars in the SMC cluster NGC\,330 ($\sim 30-40$\,Myr) and the LMC cluster NGC\,1850 ($\sim 80$\,Myr) have been conducted \citep{2020A&A...634A..51B,2021A&A...652A..70B,2023A&A...680A..32B,2023MNRAS.526..299S,2023MNRAS.518.1505K}, but the nature of evolved massive binaries is yet to be discovered. Similar observations for about ten LMC clusters, whose ages are covered by our simulations, are ongoing, with data expected soon. We plan to compare our models with these observations in detail in subsequent studies, and thereby enhance our understanding of binary evolution.

\section*{Acknowledgments}
CW is grateful to Jan Eldridge, Nikolay Britavskiy, Dietrich Baade, and Luqian Wang for their insightful discussions. We thank the anonymous referee for constructive comments.
CW, SJ, and AVG acknowledge funding from the Netherlands Organisation for Scientific Research (NWO), as part of the Vidi research program BinWaves (project number 639.042.728, PI: de Mink).
EL acknowledges funding by the European Research Council (ERC) under the European Union’s Horizon 2020 research and innovation program (Grant agreement No. 945806) and support by the Klaus Tschira Stiftung.

\balance

\bibliographystyle{aasjournal}
\bibliography{sample}{}

\clearpage
\newpage
\appendix

\section{Distribution of binary models in the color-magnitude diagram across various ages}\label{app_sec:other_CMD}
\setcounter{figure}{0}
\renewcommand\thesection{\Alph{section}}
\renewcommand{\thefigure}{\thesection.\arabic{figure}}

\begin{figure*}[htbp]
\centering
\includegraphics[width=1\linewidth]{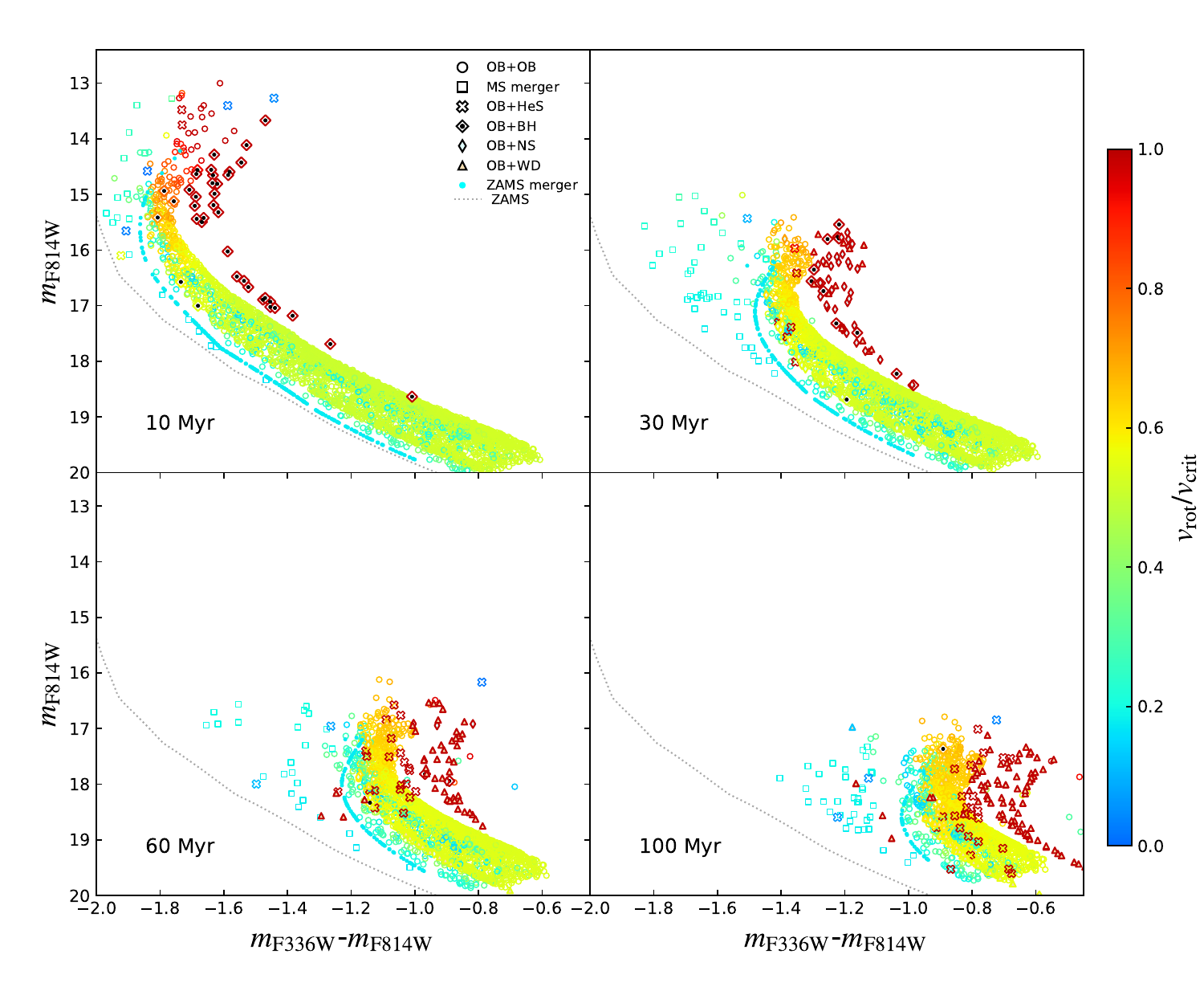}
\caption{Similar as Fig.\,\ref{fig:xxu_animation}, but for ages of 10, 30, 60 and 100\,Myr. 
}
\label{app_fig:4CMD}
\end{figure*}

Figure\,\ref{app_fig:4CMD} displays the distribution of our detailed binary models at four specific ages. These distributions match the snapshots featured in our animation, but the color and magnitude ranges in the four panels of this figure are set to be the same, such that the evolution of the cluster turn-off can be seen clearly.

\newpage
\section{Predicted properties of helium-star and compact object binaries at fixed ages}\label{app_sec:other_ages}
\setcounter{figure}{0}
\renewcommand\thesection{\Alph{section}}
\renewcommand{\thefigure}{\thesection.\arabic{figure}}

\begin{figure*}[htbp]
\centering
\includegraphics[width=1\linewidth]{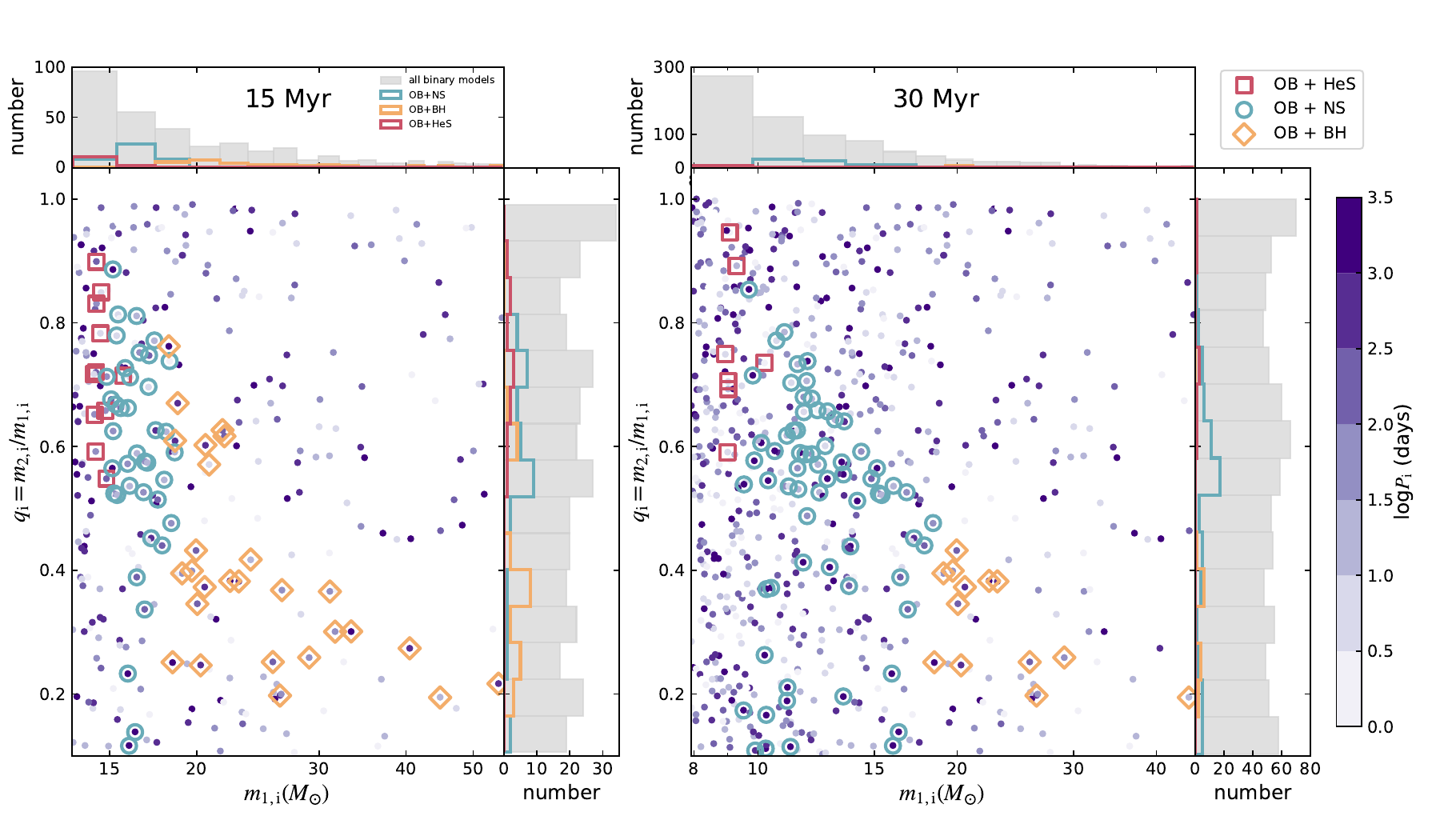}
\caption{Initial parameters for evolved massive binaries at 15 (left) and 30 Myr (right). Each individual dot represents a simulated binary model, with its color reflecting its initial orbital period. The red brown, turquoise, and orange open symbols mark the binaries currently appearing as OB+helium burning stars, OB+neutron stars (zero-kick), and OB+black hole systems, respectively. The small dots not surrounded by open symbols represent systems not identified as such evolved massive binaries (e.g. merger products, post-MS stars and pre-interaction binaries) at these ages. 
The 1D projections of the data are depicted on the upper and right margins of the primary plot.
}
\label{app_fig:init_par}
\end{figure*}

In this appendix, we extend the results shown in Fig.\,\ref{fig:BH_MS_15} to more ages. 
We begin by showing the initial binary parameters for evolved massive binaries at 15 and 30\,Myrs in Fig.\,\ref{app_fig:init_par}.
As elucidated in the main manuscript, the initial primary star mass $m_{\rm 1,i}$ of BH and NS progenitors can considerably exceed the cluster turn-off mass. In contrast, $m_{\rm 1,i}$ for HeS progenitors is generally only slightly above the cluster turn-off mass, except in case of Case A binaries. Fig\,\ref{app_fig:init_par} reveals an inverse correlation between the initial primary star mass $m_{\rm 1,i}$ and the maximum mass ratio $q_{\rm i}$ for OB+cc binaries. This inverse relationship is due to secondary stars in larger mass ratio binaries having evolved past their MS phases. Conversely, at 15 and 30\,Myr, all OB+HeS binaries exhibit initial mass ratios above roughly 0.6, as binaries with smaller ratios tend to merge during interaction phases. This suggests that in clusters older than 15\,Myr, binaries with mass ratios less than approximately 0.6 evolve into OB+cc binaries only if they are non-interactive (i.e., $\log P{\rm i} \gtrsim 3$).
The small dots not covered by open markers (i.e. not identified as evolved massive binaries) represent objects that are either merger products, post-MS stars, or pre-interaction binaries.

\begin{figure*}[htbp]
\centering
\includegraphics[width=0.8\linewidth]{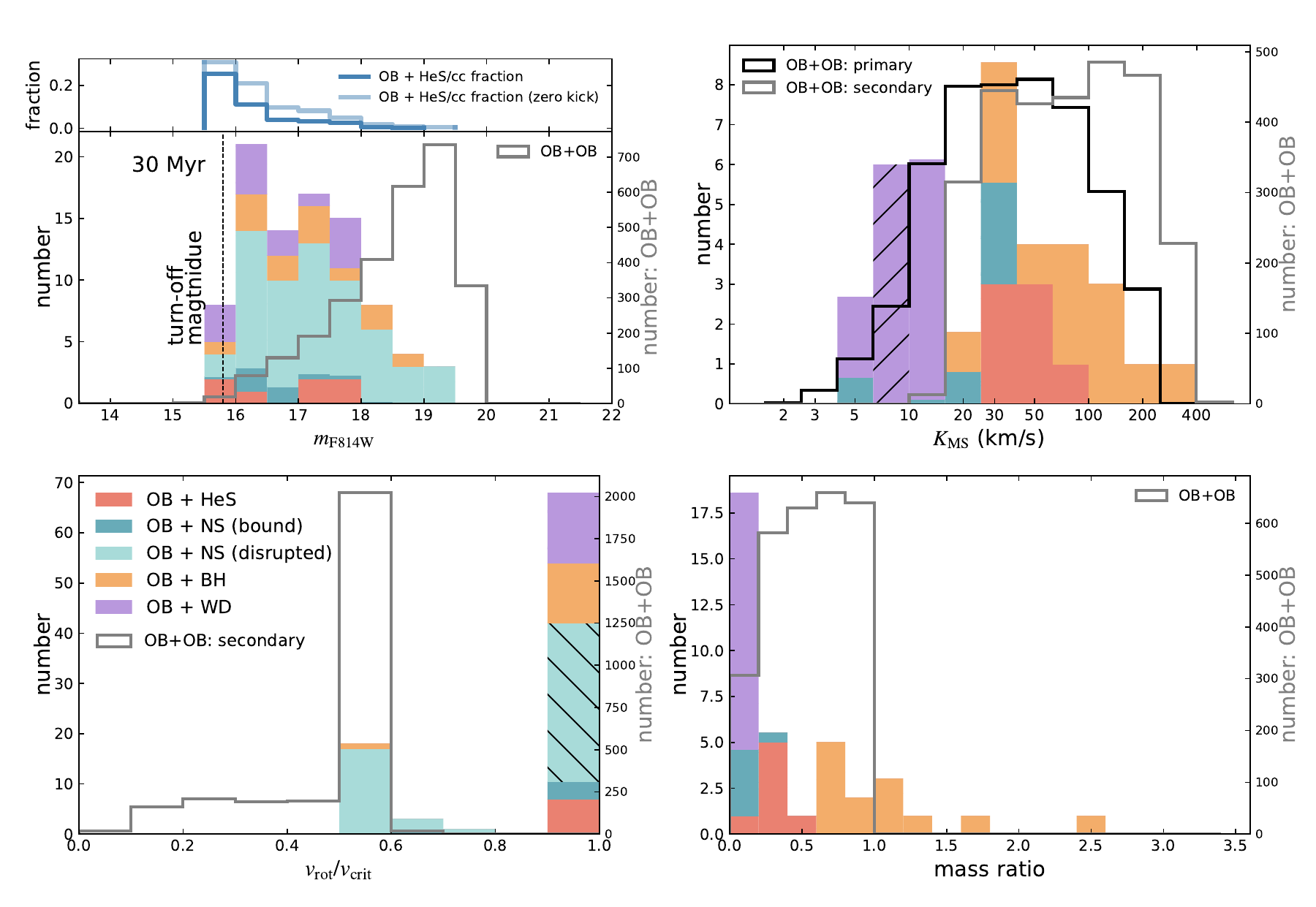}
\caption{Same as Fig.\,\ref{fig:BH_MS_15}, but for an age of 30\,Myr.
}
\label{app_fig:30Myr}
\end{figure*}

\begin{figure*}[htbp]
\centering
\includegraphics[width=0.8\linewidth]{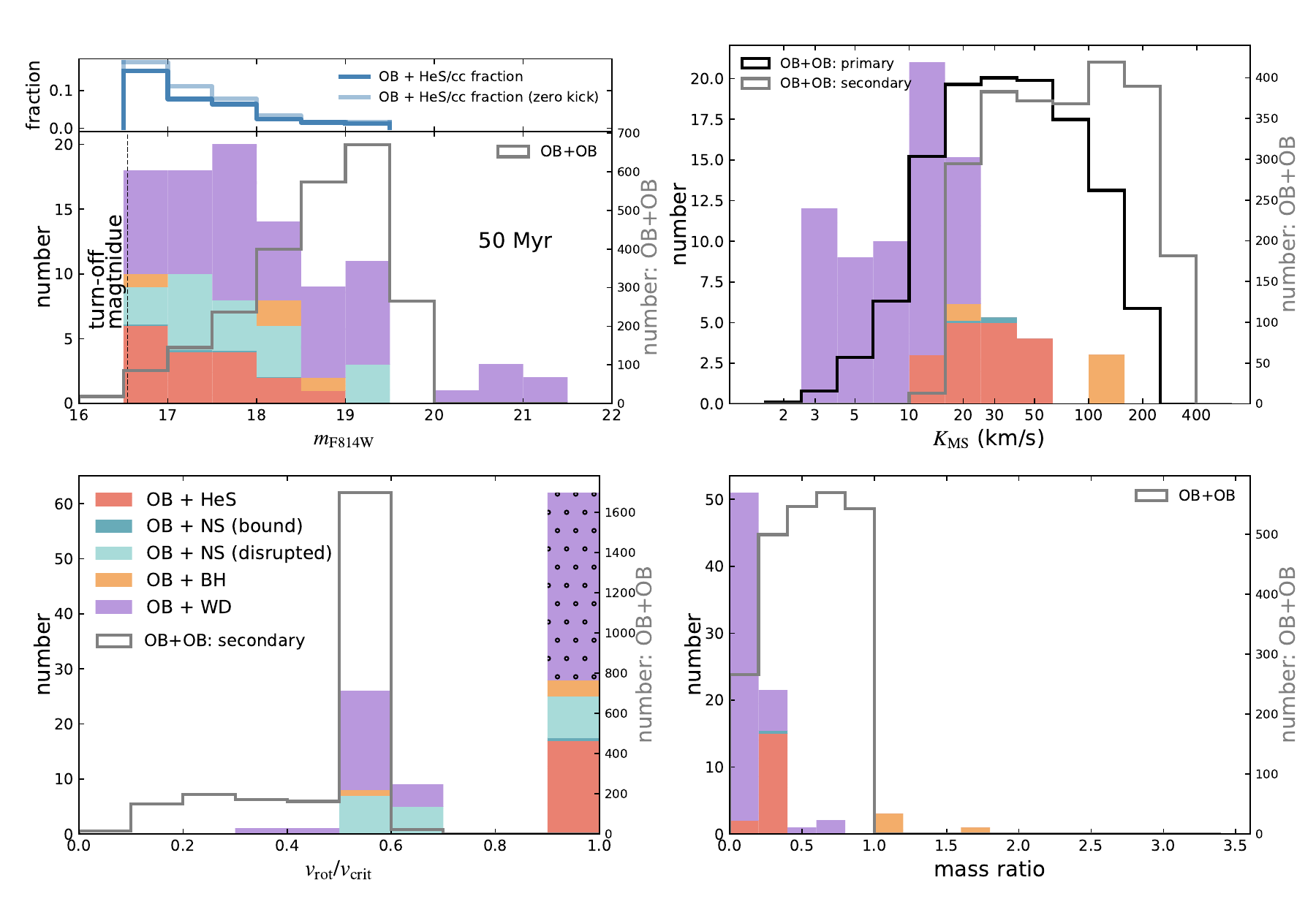}
\caption{Same as Fig.\,\ref{fig:BH_MS_15}, but for an age of 50\,Myr.
}
\label{app_fig:50Myr}
\end{figure*}

\begin{figure*}[htbp]
\centering
\includegraphics[width=0.8\linewidth]{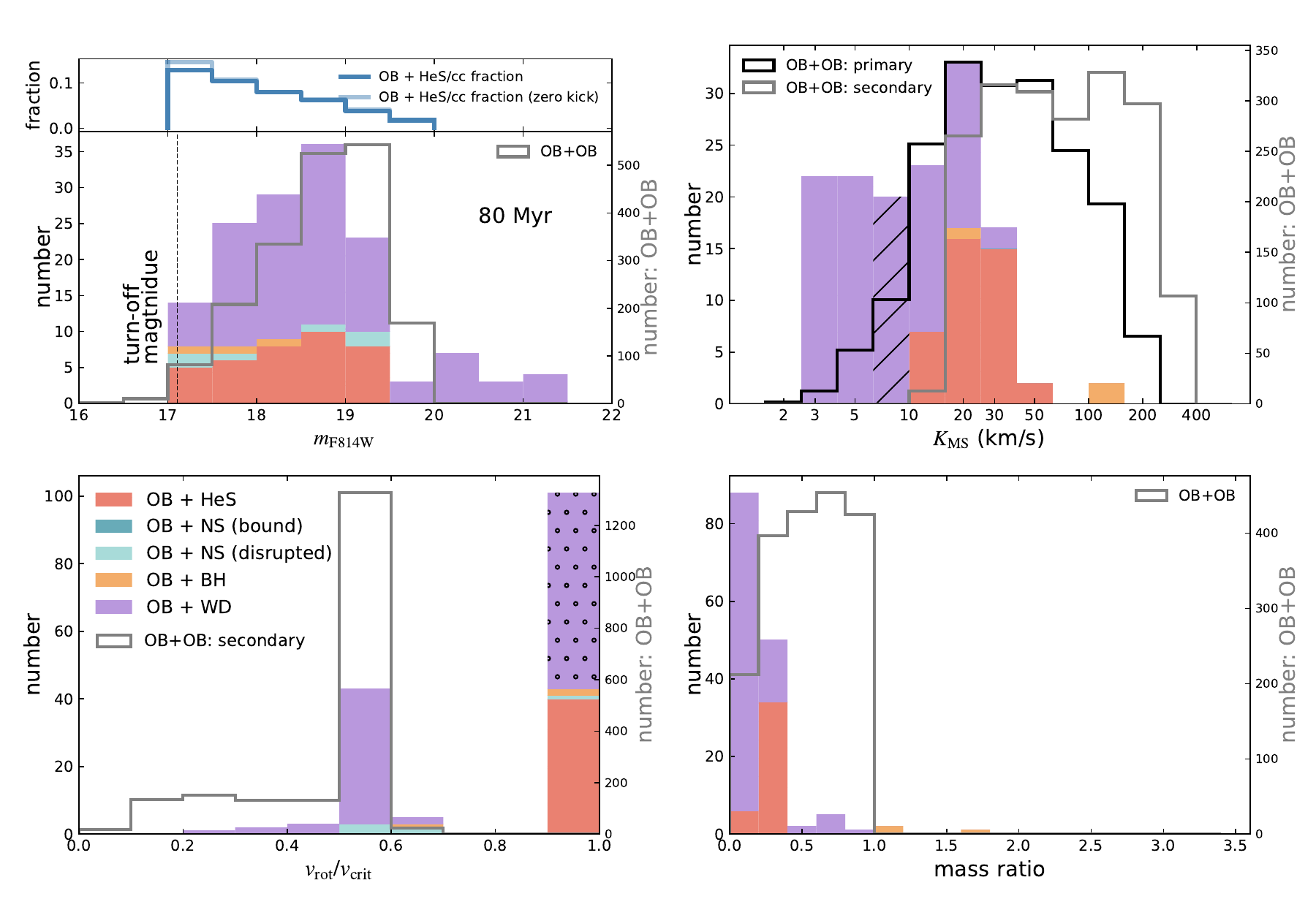}
\caption{Same as Fig.\,\ref{fig:BH_MS_15}, but for an age of 80\,Myr.
}
\label{app_fig:80Myr}
\end{figure*}

Figures\,\ref{app_fig:30Myr} to \ref{app_fig:80Myr} illustrate the distribution of evolved massive binaries at three additional ages, 30\,Myr, 50\,Myr and 80\,Myr. These coincide with the ages of the SMC cluster NGC\,330, the LMC clusters NGC\,1755, and NGC\,1850, respectively. 
Across these ages, a consistent trend emerges where the proportion of evolved binaries diminishes with increasing magnitude. 
Similar to the findings in Fig.\,\ref{fig:BH_MS_15}, OB stars in post-interaction binaries tend to have critical velocity. Conversely, those in non-interacting binaries retain their original velocities. Although close Case A binaries produce OB stars with varied velocities, their numbers remain relatively limited. 

\clearpage

\section{Predicted masses of helium-star/neutron star/white dwarf binaries}\label{app_sec:other_binaries}

\setcounter{figure}{0}
\renewcommand\thesection{\Alph{section}}
\renewcommand{\thefigure}{\thesection.\arabic{figure}}

\begin{figure*}[htbp]
\centering
\includegraphics[width=1\linewidth]{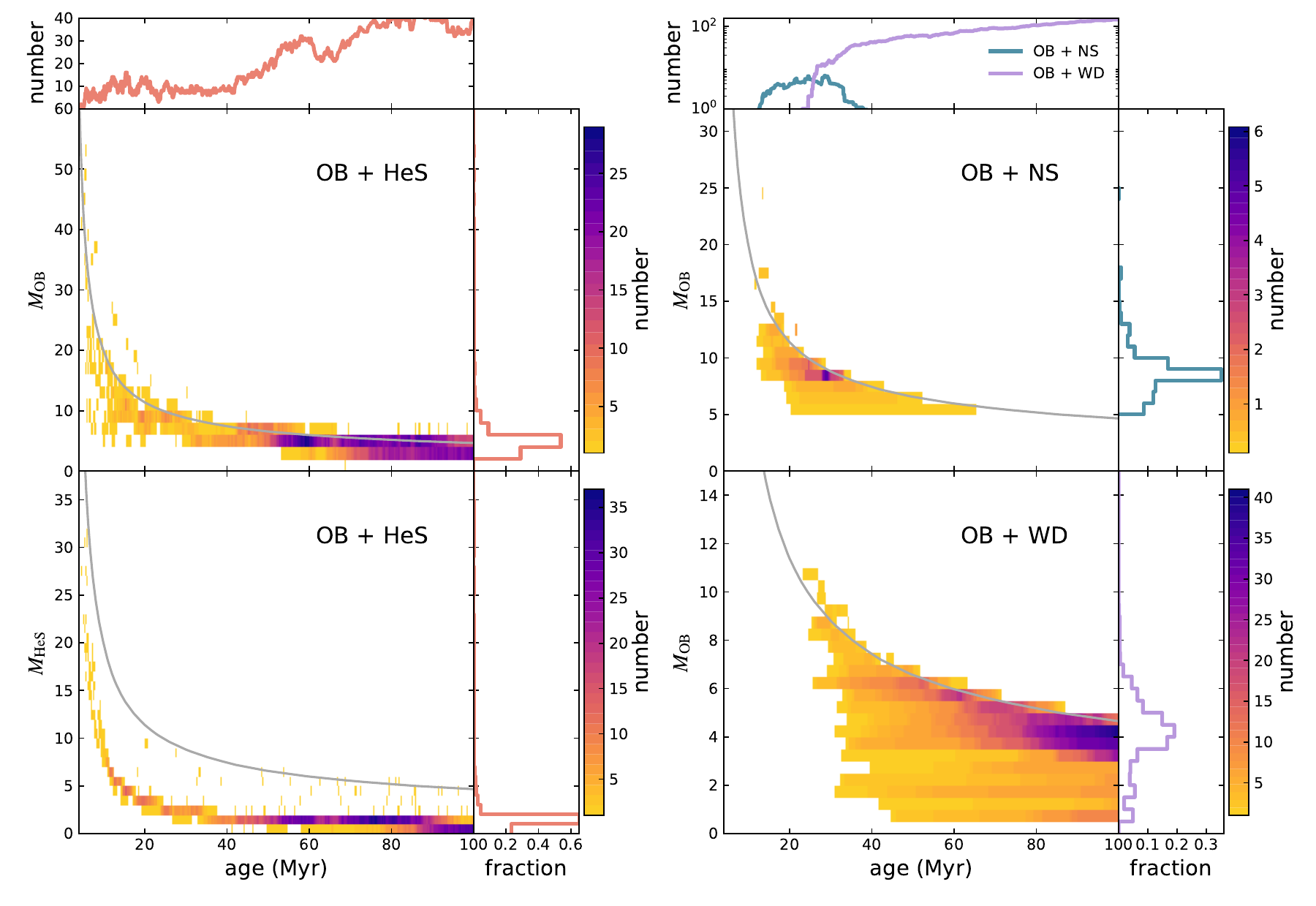}
\caption{Predicted mass distribution for binaries comprising an OB star paired with a helium-star, neutron star, or white dwarf component, similar to Fig.\,\ref{fig:BH_MS}. Colors indicate the count of respective binaries in each pixel. For OB +HeS binaries, the mass distribution is depicted for both stellar components. 
}
\label{app_fig:properties_other}
\end{figure*}

Figure\,\ref{app_fig:properties_other} depicts the predicted masses of binaries consisting of an OB star and a HeS, NS or WD. Analogous to Fig.\,\ref{fig:BH_MS}, the mass of OB stars in these binaries is typically confined by the cluster turn-off mass. Those originating from Case A mass transfer can, however, exceed this general threshold. 

The lifetime of OB+HeS binaries is defined by the short lifetime of the HeS. The well defined distribution of the mass of HeS is due to the fact that the majority of binaries are post Case B binaries, and Case B binaries with identical initial primary masses exhibit the same core masses. Nonetheless, the remaining envelope mass after mass transfer displays minor differences based on the initial orbital period. As a contrast, primary stars in Case A binaries possess smaller cores than their Case B counterparts, prolonging their MS lifetimes prior to evolving into HeSs. Consequently, HeSs emerging from Case A scenarios typically have larger masses than those from Case B at a given age. 

The rationale behind the distribution of OB star masses in OB+NS and OB+WD binaries is the same as that of the OB+BH binaries.

\newpage
\section{Effect of mass transfer stability}\label{app_sec:MT}
In this appendix, we explore the impact of uncertain mass transfer stability on the predicted counts of evolved massive binaries. As mentioned in the main manuscript, it is possible for binaries to avoid merging even if the unaccreted material cannot be effectively expelled by the radiation pressure of the two stars. For this analysis, we assume that Case B binaries, which would merge during mass transfer according to our detailed binary models, can successfully complete the mass transfer phase and produce an OB+cc binary. We refer to this assumption as ``idealized mass transfer (MT)".

We set the accretion efficiency to zero and calculate the remaining MS lifetime of the OB star using single-star models. These models have initial rotational velocities at 60\% of their critical values. Although the rotational velocity of the OB stars will increase to critical values after mass accretion, their MS lifetimes remain similar to those with 60\% critical rotation, as no mass is actually accreted by the accretor, thereby avoiding any rejuvenation.

The updated predictions for the distribution of evolved massive binaries are depicted in Fig.\,\ref{app_fig:MT_test}. Unsurprisingly, there is an increase in the count of OB+cc binaries, with OB+WD binaries seeing a surge. Under our previous assumption, binary models with lower primary masses, which are progenitors of WDs, have a higher possibility to merge. The distribution patterns and counts of OB+BH and OB+NS binaries align with our prior conclusions. The peaks coincide with previous results, albeit with a moderate increase in binary counts at these peak values. The most discernible distinction is the increasing number of OB+BH and OB+NS binaries in older clusters, stemming from binaries with smaller mass ratios. 

\setcounter{figure}{0}
\renewcommand\thesection{\Alph{section}}
\renewcommand{\thefigure}{\thesection.\arabic{figure}}

\begin{figure}[htbp]
\centering
\includegraphics[width=0.9\linewidth]{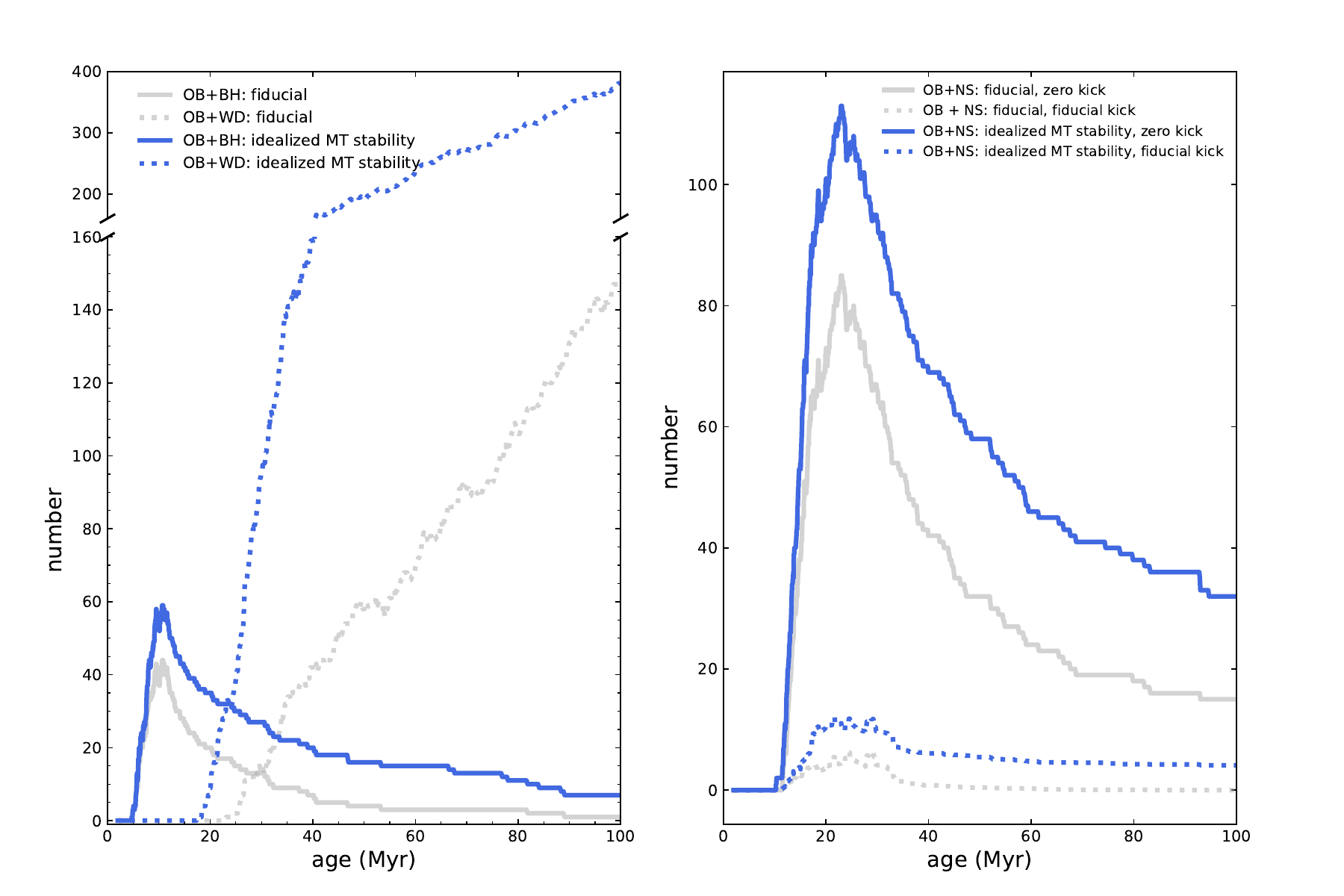}
\caption{Comparison between the number of evolved massive binaries resulting from fiducial models and assumptions of idealized mass transfer stability, in which all Case B binaries undergo stable mass transfer and produce a OB+cc binary. The left panel shows results for OB+black hole (solid lines) and OB+white dwarf binaries (dotted lines), with grey lines indicating fiducial results and blue lines indicating results based on the idealized mass transfer stability assumption. This panel is bifurcated with the major lower section retaining the same y-axis range as in Fig.\,\ref{fig:number} to facilitate a direct comparison. The right panel demonstrates the results for OB+neutron star binaries with (dotted lines) and without (solid lines) supernova kicks, again with grey lines corresponding to fiducial results and blue lines corresponding to results based on idealized mass transfer stability.  
}
\label{app_fig:MT_test}
\end{figure}

\begin{figure*}[htbp]
\centering
\includegraphics[width=1\linewidth]{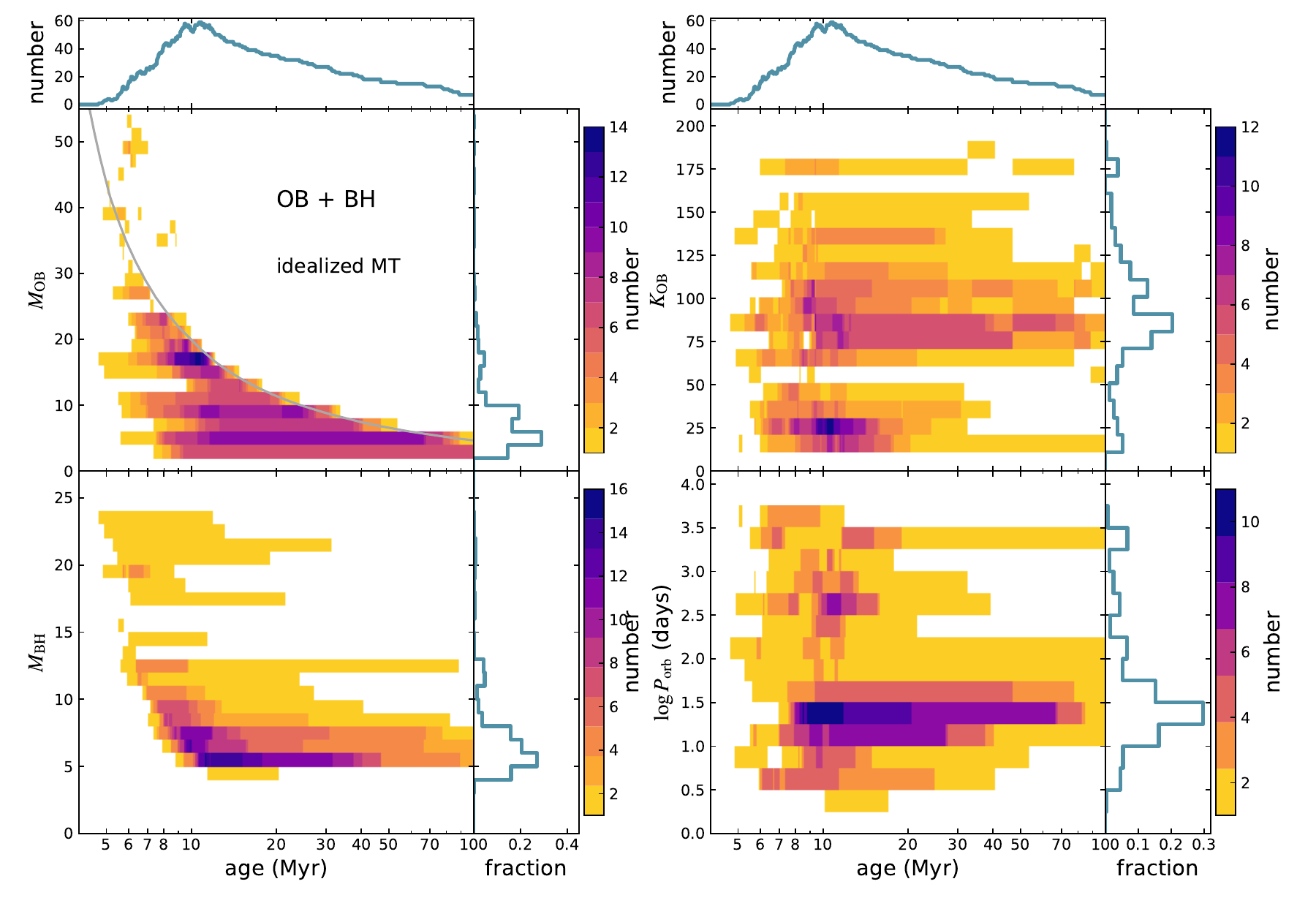}
\caption{Same as Fig.\,\ref{fig:BH_MS}, but the results are obtained with the idealized mass transfer assumption.
}
\label{app_fig:ideal_MT_properties}
\end{figure*}

\clearpage
\newpage
\section{Effect of assumptions regarding compact object formation}\label{app_sec:SN}
\setcounter{figure}{0}
\renewcommand\thesection{\Alph{section}}
\renewcommand{\thefigure}{\thesection.\arabic{figure}}
Several studies from independent groups have revealed that the final core structure of stars, in particular their final iron core masses, is non-monotonic with the initial mass \citep{1996ApJ...457..834T,2011ApJ...730...70O,2014ApJ...783...10S,2016ApJ...821...38S,2018ApJ...860...93S,2016ApJ...818..124E,2019ApJ...878...49W,2020ApJ...890...43C}. This results in a non-monotonic pattern for the formation of BHs and NS.
In this appendix, We adopt prescriptions on the formation of compact objects and their masses in \citet{2021A&A...645A...5S} derived from detailed analyses of pre-SN model of binary stripped stars. 
They found that binary stripping considerably expands the mass range for successful SN explosions compared to single star evolution, due to the reduced CO core size and the systematically different core structure (i.e. the central carbon mass fraction) at the end of helium burning \citep{2021A&A...656A..58L}. Only models with $m_{\rm 1,i}\sim 31\mso$ and $m_{\rm 1,i}\ge 65\mso$ evolve to BHs. We employ Eq.\,B.4 from \citet{2021A&A...645A...5S} to compute the remnant mass of the primary stars, which uses the CO core mass at the point of center He depletion.
For simplicity, we apply their fitting formula for Case B binaries to all our binaries, because the majority of OB+cc binaries in our models are from Case B mass transfer. BHs are identified as remnants with masses exceeding $2\mso$. The delineation between NS and WD formation remains unchanged. 

\begin{figure}[htbp]
\centering
\includegraphics[width=0.7\linewidth]{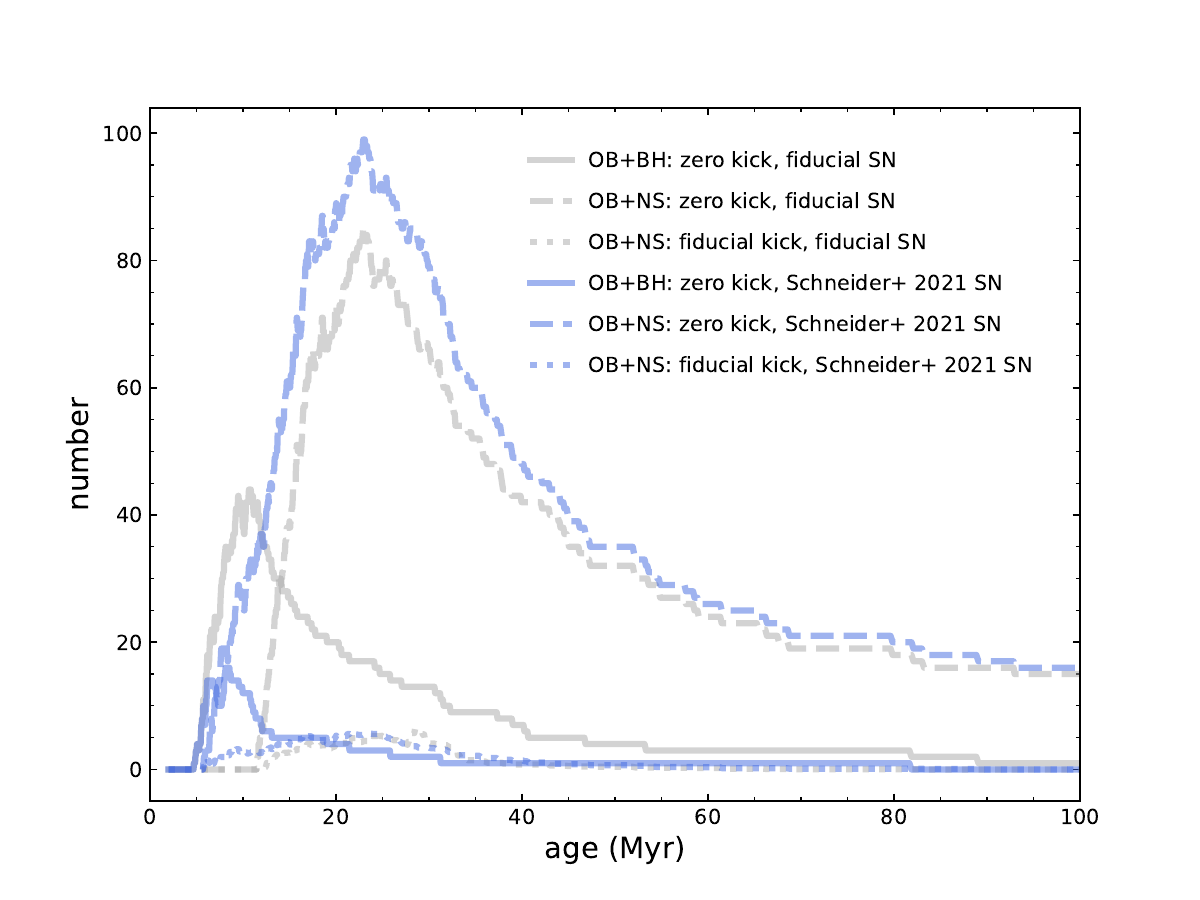}
\caption{Comparison between the number of evolved massive binaries resulting from fiducial models (grey lines) and the assumptions regarding supernova explosions proposed by \citet{2021A&A...645A...5S} (blue lines). The solid lines show the results for OB+black hole binaries. The dashed and dotted lines show the results for OB+neutron star binaries without and with supernova kicks, respectively. 
}
\label{app_fig:SN_Fabian}
\end{figure}

The results based on this new assumption are illustrated in Fig.\,\ref{app_fig:SN_Fabian}. We see a noticeable decline in the number of OB+BH binaries, due to the significant reduction in parameter space for BH formation. While the number of progenitors of OB+NS binaries rises, many of these systems are disrupted by SN kicks. As a consequence, the number of OB+NS binaries does not increase significantly, but these binaries appear in younger clusters. 



\begin{figure*}[htbp]
\centering
\includegraphics[width=0.9\linewidth]{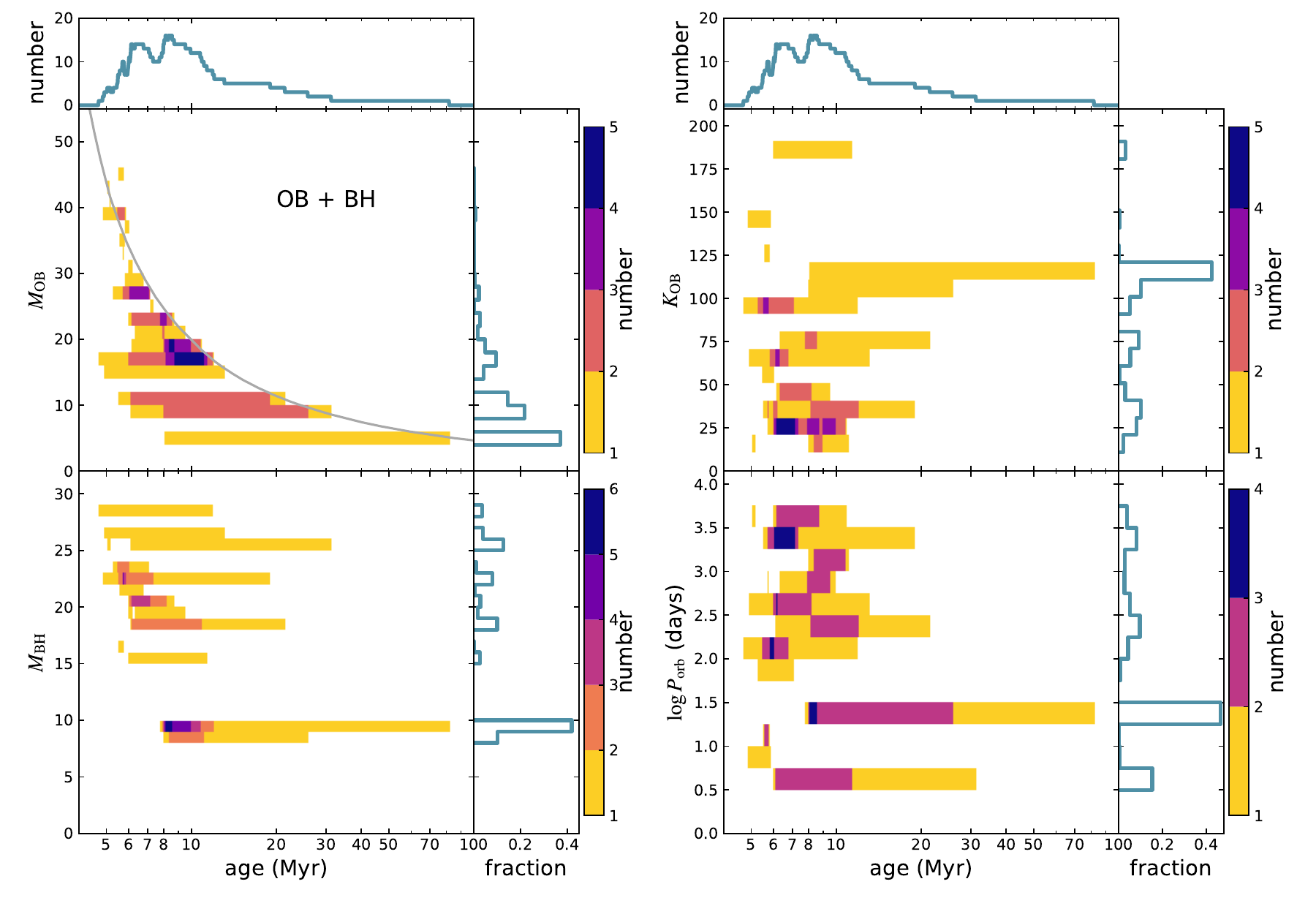}
\caption{Same as Fig.\,\ref{fig:BH_MS}, but the results are based on an alternative supernova explosion assumption as proposed by \citet{2021A&A...645A...5S}.
}
\label{app_fig:SN_Fabian_properties}
\end{figure*}


\clearpage
\newpage
\section{Effect of sample size}\label{app_sec:size}
In this appendix, we explore how varying the sample size affects the reliability of our predictions for the number of evolved massive binaries. We systematically reduce our binary model sample to 60\%, 40\%, 20\%, and 10\% of the original size and present the results in Figure\,\ref{app_fig:reduced_mass}. The pattern of predicted numbers remains stable even when the sample is reduced to 10\%, corresponding to a cluster mass of approximately $1.3\times 10^4 \mso$. Young open clusters in the Magellanic Clouds typically have total stellar masses ranging from a few thousand to several tens of thousands of solar masses \citep{2003AJ....126.1836H,2012ApJ...751..122P}, with those displaying distinct multiple MSs often exceeding tens of thousands of solar masses \citep{2018MNRAS.477.2640M}. Consequently, our simulations are particularly appropriate for studying young open clusters with multiple MS components.

\setcounter{figure}{0}
\renewcommand\thesection{\Alph{section}}
\renewcommand{\thefigure}{\thesection.\arabic{figure}}

\begin{figure*}[htbp]
\centering
\includegraphics[width=1\linewidth]{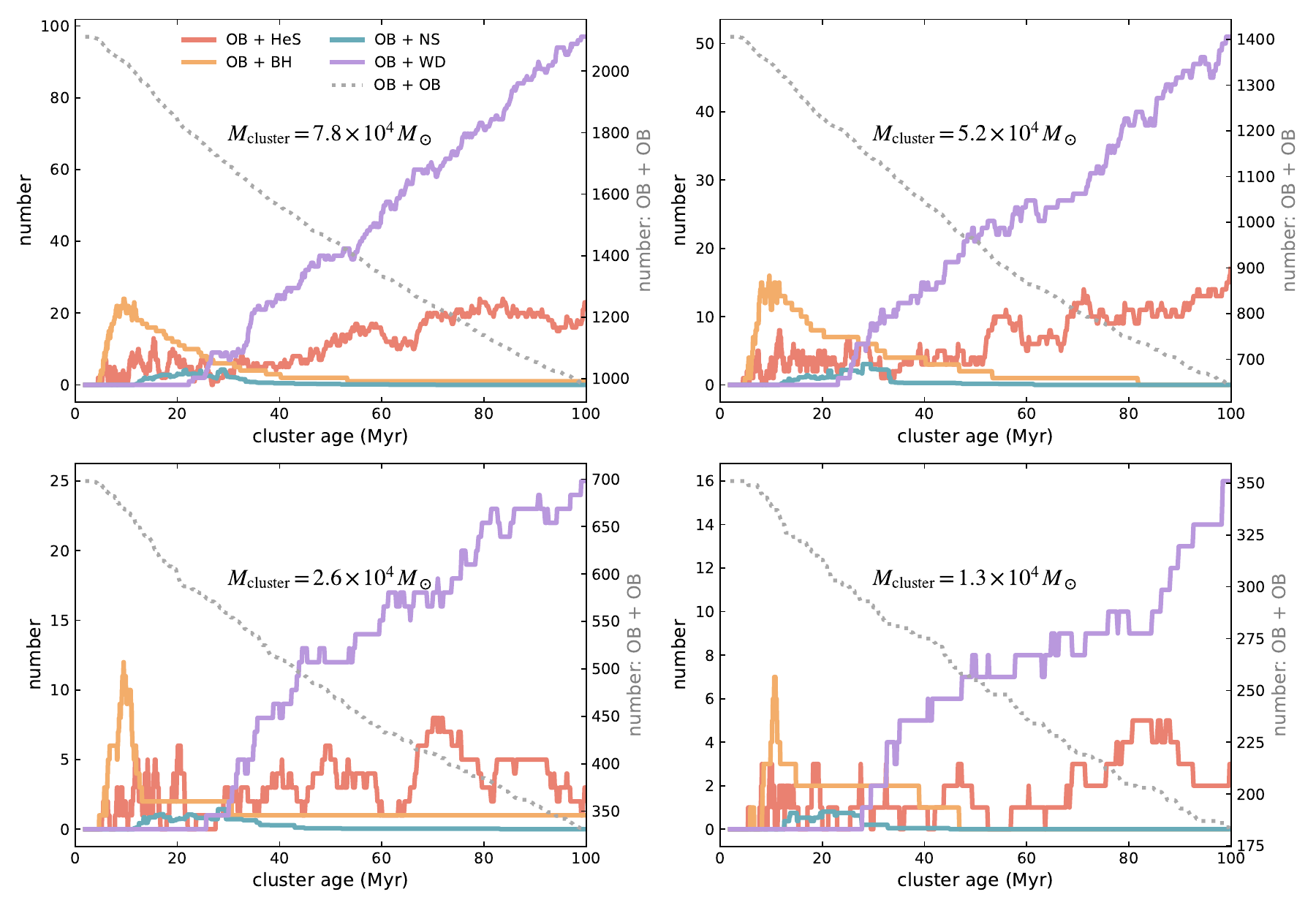}
\caption{Same as Fig.\,\ref{fig:number}, but the sample size is reduced to 60\%, 40\%, 20\% and 10\% in each panel.
}
\label{app_fig:reduced_mass}
\end{figure*}





\end{document}